\documentclass[12pt]{article}
\usepackage{graphics}
\usepackage{graphicx}
\DeclareGraphicsExtensions{.pdf}
\usepackage{float} 
\usepackage{amsmath}
\usepackage{slashed}
\usepackage{amsfonts}   
\usepackage{amssymb}

\def\c{\cite}
\usepackage{color}

\begin{document}
\date{}
\title{{\bf{\Large Nonrelativistic expansion of Green-Schwarz strings on $ AdS_5 \times S^5 $}}}
\author{
 {\bf {\normalsize Dibakar Roychowdhury}$
$\thanks{E-mail:  dibakarphys@gmail.com, dibakar.roychowdhury@ph.iitr.ac.in}}\\
 {\normalsize  Department of Physics, Indian Institute of Technology Roorkee,}\\
  {\normalsize Roorkee 247667, Uttarakhand, India}
\\[0.3cm]
}

\maketitle
\begin{abstract}
We perform nonrelativistic expansions of type IIB $ AdS_5 \times S^5 $ Green-Schwarz (GS) superstrings upto quadratic order in fermionic excitation. We carry out a systematic $ 1/c $ expansion both for the bosonic as well as the fermionic sector of the type IIB model and explore the LO theory in detail. We explore various global as well as local symmetries of the nonrelativistic sigma model at LO including the world-sheet supersymmetry. As a special case, we choose the flat gauge and explore the longitudinal T-duality rules for the LO theory. Finally, we comment on the NLO theory and discuss some of its interesting properties including the scale invariance.
\end{abstract}
\section{Overview and general motivation of the paper}
Nonrelativistic (NR) string theory, that was originally pioneered due to the seminal work of Gomis and Ooguri \cite{Gomis:2000bd}, has been subsequently extended and probed into deeper in the recent years. Starting with some earlier developments \cite{Danielsson:2000gi}-\cite{Gomis:2016zur}, we now have a clear understanding about the target space of NR string theory \cite{Bergshoeff:2018yvt}-\cite{Yan:2021lbe} and brane like objects \cite{Ebert:2021mfu}-\cite{Roychowdhury:2022est} in general.

There have been two parallel and mutually compatible \cite{Harmark:2019upf} approaches to NR strings. While considering the String Newton Cartan (SNC) limit \cite{Bergshoeff:2018yvt} of relativistic strings, one typically singles out one time and one space direction (those are termed as the longitudinal axes of the resulting Non-Lorentzian (NL) target space) from the rest of the target space directions called the transverse axes. In other words, this is a co-dimension two foliation approach to NR strings which enjoys all the global symmetries of the associated NL background. The second approach goes under the name of the null reduction of Lorentzian backgrounds where the NR strings are coupled to NL target space that is obtained via dimensional reduction along a null isometry of the original relativistic target space \cite{Harmark:2017rpg}-\cite{Harmark:2018cdl}.

The theme of the present paper is however based on the recently proposed string $1/c^2$ expansion \cite{Hartong:2021ekg}-\cite{Hartong:2022dsx} of the world-sheet bosonic degrees of freedom while the NR string is coupled to a NL background that is characterized by a co-dimension two foliation as mentioned above. The question that we pose here, what if and whether this formalism could be extended in the presence of fermionic excitations. We address this issue taking a specific example of the Green-Schwarz (GS) superstrings \cite{Green:1983wt}-\cite{Green:1983sg} on $ AdS_5 \times S^5 $ \cite{Metsaev:1998it}-\cite{Bellucci:2005vq} and subsequently carrying out a NR expansion of that. 

During the process of NR reduction, the relativistic sigma model eventually boils down into a covariant action that can be systematically expressed as a string $ 1/c $ expansion\footnote{In the presence of fermions, the NR sigma model can be typically expanded as, $S_{NR}\sim S^{(LO)}_{NR} + c^{-1}S^{(NLO)}_{NR}+ \mathcal{O}(c^{-2})$. The LO theory, as we see, receives contributions that are coming both from the bosonic as well as the fermionic sectors of the parent GS sigma model. On the other hand, the NLO contributions that appear at $\mathcal{O}(c^{-1})$ are purely due to the fermionic quadratic terms present in the parent GS action \cite{Metsaev:1998it}-\cite{Bellucci:2005vq}. To summarize, we therefore denote this overall NR expansion as a string $1/c$ expansion.}. The corresponding bosonic sector of the Non-Lorentzian (NL) target space is identified as the String Newton-Cartan (SNC) background which is characterized by the set of (SNC) data \cite{Bergshoeff:2018yvt} those are introduced and explained in detail in Section \ref{sec2}.

We carry out an analogous $ 1/c $ expansion for the fermionic sector in Section \ref{sec3}, where we expand the world-sheet spin connection and the fermion mass term \cite{Hernandez:2004kr}-\cite{Bellucci:2005vq} that appears at quadratic order in fermions. These expansions follow directly by considering a co-dimension two foliation of the target space accompanied by a $ 1/c^2 $ expansion \cite{Hartong:2021ekg} for the bosonic fields.

We figure out the NR sigma model order by order in a string $ 1/c $ expansion and in particular focus on the LO and NLO Lagrangian densities. We discuss the LO theory in detail in Section \ref{sec4}.  As mentioned before, the LO theory receives contributions both from the bosonic and the fermionic sectors of the parent GS model and turns out to be the simplest NR theory to be worked on. We investigate various global as well as local symmetries of the LO theory including the local supersymmetry \cite{Gomis:2016zur}, \cite{Hyun:2000hr}. We show that the classical 2d stress tensor is traceless which reflects the Weyl invariance of the NR sigma model at LO. 

We further use these symmetries to gauge fix the LO theory. Using the conformal gauge, we further probe into Galilean boost symmetries of the LO theory and derive the associated transformation rules for the bosonic as well as the fermionic degrees of freedom. We show that for the fermionic sector, the boost invariance is realized subjected to certain specific transformation rules pertaining to the world-sheet fermions and their derivatives. 

We also discuss the dilatation and/or the (local) scale invariance \cite{Blair:2021waq}-\cite{Roychowdhury:2022est} of the LO theory. It turns out that in order to preserve the scale invariance at LO, the fermions and their derivatives must transform in a non-trivial fashion. Down the line, we perform (local) supersymmetry variation of the NR Lagrangian and identify the supersymmetry transformation rules for the fermionic and bosonic degrees of freedom at LO.

Next, we consider special case - namely the choice of the \emph{flat} gauge and explore the T-duality transformation rules \cite{Rocek:1991ps} for the LO theory within this gauge choice. The isometry direction that we T-dualize, happens to be the longitudinal spatial axes of the NL target space. In other words, the LO theory allows us to perform a longitudinal T-duality \cite{Bergshoeff:2018yvt}.

Finally, we conclude in Section \ref{sec5}, where we outline some key features of the NLO theory. In particular, we show that the Weyl invariance of the NR sigma model is preserved at that order. A closer look further reveals that the field transformations and their solutions at NLO are governed by the solutions at the preceding order and this turns out to be the generic feature which continues to hold at subsequent orders in the NR expansion. 
\section{Bosonic sector}
\label{sec2}
The purpose of this section is to construct the NR GS sigma model action following a string $ 1/c $ expansion \cite{Hartong:2021ekg}-\cite{Hartong:2022dsx}. We consider that the string is propagating over $ AdS_5 \times S^5 $
\begin{align}
\label{e2.1}
&ds_{AdS_5}^2 = L^2[-\cosh^2 \rho dt^2 + d\rho^2 + \sinh^2\rho (d\theta^2 + \cos^2\theta d\phi^2_1 + \sin^2\theta d\phi^2_2)],\\
&ds^2_{S^5}=L^2 [d \zeta^2 +\cos^2 \zeta d \varphi^2_3 + \sin^2\zeta (d\psi^2 + \cos^2\psi d \varphi^2_1 +\sin^2\psi d\varphi^2_2)],
\label{e2.2}
\end{align}
which is supported by a self dual RR five form flux\footnote{We scale $ L=c \ell $ while considering NR expansion of the metric.}
\begin{align}
F_5 =\frac{4}{L}(\text{Vol}(AdS_5)-\text{Vol}(S^5)).
\end{align}

We propose the following $ 1/c $ expansion for the bosonic target space vielbeins \cite{Bergshoeff:2018vfn}-\cite{Bergshoeff:2021bmc}
\begin{align}
\label{e2.3}
&E_t~^A = c \tau_t~^A +c^{-1}m_t~^A\\
&E_\rho~^A = c \tau_\rho~^A +c^{-1}m_\rho~^A,
\label{2.5}
\end{align}
where $\lbrace t, \rho \rbrace$ correspond to \emph{longitudinal} directions of the Non-Lorentzian (NL) manifold.

The inverse vielbeins are defined as
\begin{align}
&E^t~_A= c^{-1}\tau^t~_A + c^{-3}m^t~_A\\
& E^\rho~_A =c^{-1} \tau^\rho~_A + c^{-3}m^\rho~_A.
\label{2.7}
\end{align}

Following $ E_t~^A  E^t~_B = \delta^A_B$, we find
\begin{align}
\tau_t~^A \tau^t~_B = \delta^A_B ~;~ \tau_t~^A m^t~_B + m_t~^A \tau^t~_B=0.
\end{align}

On the other hand, the \emph{transverse} vielbeins can be expressed as\footnote{Here, $ X^{\bar{a}} =t, \rho$ correspond to the longitudinal axes of the NL target space.}
\begin{align}
\label{e2.5}
E_{\hat{\mu}}~^I =L\sqrt{f(X^{\bar{a}})} e_{\hat{\mu}}~^I,
\end{align}
where $ \hat{\mu}$ collectively denotes the rest of the transverse directions of the NL target space.

The corresponding inverse vielbein is defined as
\begin{align}
E^{\hat{\mu}}~_I =\frac{L^{-1}}{\sqrt{f(X^{\bar{a}})}}e^{\hat{\mu}}~_I,
\end{align}
such that one satisfies the normalisation condition
\begin{align}
E^{\hat{\mu}}~_J E_{\hat{\mu}}~^I =\delta^I_J =e^{\hat{\mu}}~_J e_{\hat{\mu}}~^I.
\label{2.11}
\end{align}

As it stands, under NR scaling \eqref{e2.8} one could expand
\begin{align}
f(X^{\bar{a}})= f(x^{\bar{a}})+c^{-2}y^{\bar{a}}\partial_{\bar{a}}f + \mathcal{O}(c^{-4}).
\end{align}

Furthermore, we demand that under NR expansion
\begin{align}
e_{\hat{\mu}}~^I (X)= c^{-1}\tilde{e}_{\hat{\mu}}~^I(\tilde{X})~;~e^{\hat{\mu}}~_I(X)= c~ \tilde{e}^{\hat{\mu}}~_I (\tilde{X}),
\end{align}
where $ X $ and $ \tilde{X} $ are related via suitable $ 1/c $ scaling of the form $ X = c^{-1}\tilde{X} $. The $ 1/c $ scaling combined with the $ L= c \ell $ scaling keeps the volume of the transverse space finite.

Under NR scaling and using the above set of expansions \eqref{e2.3}-\eqref{e2.5}, it is trivial to show
\begin{align}
\label{2.9}
&ds^2 = c^2 T_{\bar{a}\bar{b}}(X)dX^{\bar{a}}dX^{\bar{b}}+\Pi_{\bar{a}\bar{b}}(X)dX^{\bar{a}}dX^{\bar{b}}\nonumber\\&+\ell^2 f(x^{\bar{a}})H_{\hat{\mu}\hat{\nu}}(\tilde{X})d\tilde{X}^{\hat{\mu}}d\tilde{X}^{\hat{\nu}}+\mathcal{O}(c^{-2}).
\end{align}

Notice that, here each of the above functions $ \tilde{e}_{\hat{\mu}}~^I(\tilde{X}) $ can be further expanded following a $ 1/c^2 $ expansion\footnote{Following the notational simplicity, we would not use tildes any more.} \eqref{e2.8}
\begin{align}
\label{2.10}
\tilde{e}_{\hat{\mu}}~^I(\tilde{X}) =\tilde{e}_{\hat{\mu}}~^I(\tilde{x}) +c^{-2} \tilde{y}^M \partial_M \tilde{e}_{\hat{\mu}}~^I(\tilde{x})+\mathcal{O}(c^{-2}).
\end{align}

We denote the above functions in \eqref{2.9} as
\begin{align}
\label{e2.7}
T_{\bar{a}\bar{b}}=\eta_{AB}\tau_{\bar{a}}~^A \tau_{\bar{b}}~^B~;~\Pi_{\bar{a}\bar{b}}=\eta_{AB}(\tau_{\bar{a}}~^A m_{\bar{b}}~^B +m_{\bar{a}}~^A \tau_{\bar{b}}~^B)~;~H_{\hat{\mu}\hat{\nu}}=\delta_{IJ}\tilde{e}_{\hat{\mu}}~^I \tilde{e}_{\hat{\nu}}~^J,
\end{align}
together with $ \eta_{AB}=\text{diag}(-1,1) $ being the two dimensional Lorentzian metric. As we show below, $ H_{\hat{\mu}\hat{\nu}} $ is the collective contribution due to the transverse metric components that is produced through $ S^3 \subset AdS_5 $ and the internal space $ S^5 $.
\subsection{Expansion of the background data}
Given the metric \eqref{e2.3}, it is trivial to decode the NR data associated with the NL target space. In order to extract longitudinal data, we propose the following $ 1/c^2 $ expansion \cite{Hartong:2022dsx}
\begin{align}
\label{e2.8}
X^M = x^M + c^{-2}y^M +\mathcal{O}(c^{-4})~,~\lbrace X^M= X^{\bar{a}},X^{\hat{\mu}}\rbrace.
\end{align}

This leads to the following expansions
\begin{align}
\label{e2.9}
&T_{\bar{a}\bar{b}}=\eta_{AB}\tau_{\bar{a}}~^A (x) \tau_{\bar{b}}~^B (x)+c^{-2}\eta_{AB}\Big[ \tau_{\bar{a}}~^A (x)y^M \partial_M \tau_{\bar{b}}~^B (x)\nonumber\\
&+\tau_{\bar{b}}~^B (x)y^M \partial_M \tau_{\bar{a}}~^A (x) \Big]+\mathcal{O}(c^{-4})\nonumber\\
&=T^{(0)}_{\bar{a}\bar{b}}(x)+c^{-2}T_{\bar{a}\bar{b}}^{(-2)}(x,y)+\mathcal{O}(c^{-4}),\\
\label{e2.10}
&\Pi_{\bar{a}\bar{b}}=\eta_{AB}\Big[ \tau_{\bar{a}}~^A (x) m_{\bar{b}}~^B (x) +m_{\bar{a}}~^A (x) \tau_{\bar{b}}~^B (x) \Big]\nonumber\\&+c^{-2}\eta_{AB}\Big[ y^M \partial_M \tau_{\bar{a}}~^A (x) m_{\bar{b}}~^B(x)+y^M \partial_M m_{\bar{a}}~^A (x) \tau_{\bar{b}}~^B (x) \nonumber\\&+\tau_{\bar{a}}~^A(x)y^M \partial_M m_{\bar{b}}~^B (x) +m_{\bar{a}}~^A (x) y^M \partial_M \tau_{\bar{b}}~^B (x)\Big]+\mathcal{O}(c^{-4})\nonumber\\
&=\Pi^{(0)}_{\bar{a}\bar{b}}(x)+c^{-2}\Pi_{\bar{a}\bar{b}}^{(-2)}(x,y)+\mathcal{O}(c^{-4}),\\
&H_{\hat{\mu}\hat{\nu}}=H_{\hat{\mu}\hat{\nu}}(x)+\mathcal{O}(c^{-2})~;~ H_{\hat{\mu}\hat{\nu}}(x)=\delta_{IJ}\tilde{e}_{\hat{\mu}}~^I(x) \tilde{e}_{\hat{\mu}}~^J(x).
\end{align}

\paragraph{Decoding data for $ AdS_5 \times S^5 $:} Let us now work out the SNC data for $ AdS_5 \times S^5 $ background \eqref{e2.1}-\eqref{e2.2} that we are working with.  Using expansions \eqref{e2.9}-\eqref{e2.10}, the longitudinal component of the metric can be expressed as
\begin{align}
\label{e2.12}
&ds^2_{\parallel}=c^2 T^{(0)}_{\bar{a}\bar{b}}(x)dx^{\bar{a}}dx^{\bar{b}}+\Big[ T^{(-2)}_{\bar{a}\bar{b}}(x,y) +\Pi^{(0)}_{\bar{a}\bar{b}}(x)\Big]dx^{\bar{a}}dx^{\bar{b}}\nonumber\\
&+T^{(0)}_{\bar{a}\bar{b}}(x)(dx^{\bar{a}}dy^{\bar{b}}+dy^{\bar{a}}dx^{\bar{a}})+\mathcal{O}(c^{-2}).
\end{align}

On the other hand, following an expansion of the form
\begin{align}
& \rho = \rho^{(0)}+c^{-2}\rho^{(-2)}+\mathcal{O}(c^{-4}),\\
&t=t^{(0)}+c^{-2}t^{(-2)}+\mathcal{O}(c^{-4}),
\end{align}
it is trivial to see from \eqref{e2.1} and \eqref{e2.2}
\begin{align}
\label{e2.24}
&ds^2 = ds^2_{\parallel}+ f(\rho^{(0)})(ds^2_{\perp})_{(I)}+(ds^2_{\perp})_{(II)},\\
& f(\rho^{(0)})=\ell^2 \sinh^2\rho^{(0)}.
\end{align}

We identify the parallel components associated with the co-dimension two foliation as
\begin{align}
\label{e2.15}
&ds^2_{\parallel}=c^2 \ell^2 \Big( - (dt^{(0)})^2 \cosh^2\rho^{(0)}  +(d \rho^{(0)})^2  \Big)-2\ell^2 (dt^{(0)})^2 \rho^{(-2)}\sinh\rho^{(0)}\cosh\rho^{(0)}\nonumber\\
&+ 2\ell^2 d\rho^{(0)}d \rho^{(-2)}-2 \ell^2 dt^{(0)}dt^{(-2)}\cosh^2\rho^{(0)}+\mathcal{O}(c^{-2}).
\end{align}

Comparing \eqref{e2.12} and \eqref{e2.15}, one finds\footnote{We set $ \ell=1 $ for the rest of our analysis.}
\begin{align}
\label{2.27}
&\tau_t~^{0}= \cosh\rho^{(0)}~;~\tau_{\rho}^1 = 1,\\
\label{2.23}
& \tau_t~^{0} y^M \partial_M \tau_t~^{0} + m_t~^{0}y^M \partial_M \tau_t~^{0} +\tau_t~^{0} y^M \partial_M m_t~^0 \nonumber\\
&= \rho^{(-2)}\sinh\rho^{(0)}\cosh\rho^{(0)},\\
&T_{\rho \rho}^{(-2)}+\Pi_{\rho \rho}^{(0)}=0.
\end{align}

One can further simplify \eqref{2.23} to find
\begin{align}
\cosh \rho^{(0)}\partial_{\rho^{(0)}}m_t^0 + \sinh \rho^{(0)} m_t^0 = 0,
\end{align}
which yields a solution of the form
\begin{align}
\label{2.31}
m_t^0 = \text{sech}\rho^{(0)}.
\end{align}

The transverse metric component, on the other hand, can be expressed into two parts. The first part comes following a $ 1/c $ expansion of $ S^3 \subset AdS_5 $ 
\begin{align}
&(ds^2_{\perp})_{(I)}=H^{(I)}_{\hat{\mu}\hat{\nu}}(x)dx^{\hat{\mu}}dx^{\hat{\nu}}=d\tilde{\theta}^{2(0)} + d\tilde{\phi}^{2(0)}_1 + \tilde{\theta}^{2(0)} d\tilde{\phi}^{2(0)}_2 + \mathcal{O}(c^{-2}),
\end{align}
where each of the above fields are the leading order contribution in the expansion \eqref{2.10}.

On a similar note, the second part comes through the $ 1/c $ expansion of $ S^5$
\begin{align}
&(ds^2_{\perp})_{(II)}=H^{(II)}_{\hat{\mu}\hat{\nu}}(x)dx^{\hat{\mu}}dx^{\hat{\nu}}\nonumber\\
&=d\tilde{\zeta}^{2(0)} + d\tilde{ \varphi}_3^{2(0)} +\tilde{\zeta}^{2(0)} (d\tilde{\psi}_1^{2(0)} + \cos^2\tilde{\psi}^{(0)} d\tilde{\varphi}^{2(0)}_1 +\sin^2\tilde{\psi}^{(0)} d\tilde{\varphi}^{2(0)}_2).
\end{align} 
\subsection{Bosonic Lagrangian in $ 1/c $ expansion}
Bosonic part of the full GS action \cite{Metsaev:1998it}-\cite{Drukker:2000ep} can be expressed as
\begin{align}
&L_B =-\frac{1}{2} \sqrt{-\gamma}\gamma^{\alpha \beta}E_\alpha~^{\bar{A}}E_\beta~^{\bar{B}}\eta_{\bar{A}\bar{B}}\nonumber\\
&=-\frac{1}{2} \sqrt{-\gamma}\gamma^{\alpha \beta}\Big[(c^2 T_{\bar{a}\bar{b}} (X)+\Pi_{\bar{a}\bar{b}}(X))\partial_\alpha X^{\bar{a}}\partial_\beta X^{\bar{b}}\nonumber\\
&+f(x^{\bar{a}})H_{\hat{\mu}\hat{\nu}}(X)\partial_\alpha X^{\hat{\mu}}\partial_\beta X^{\hat{\nu}}+\mathcal{O}(c^{-2}) \Big],
\end{align}
where $ \bar{A}=(A, I) $ stands for the 10d flat indices together with $ E_\alpha~^{\bar{A}}=E_M~^{\bar{A}}\partial_\alpha X^M $.

Combining NR expansions for the world-sheet metric \cite{Hartong:2022dsx}
\begin{align}
&\gamma^{\alpha \beta}=\gamma^{\alpha \beta}_{(0)}-c^{-2}\gamma^{\alpha \beta}_{(2)}+\mathcal{O}(c^{-4}),\\
&\sqrt{-\gamma}=\sqrt{-\gamma_{(0)}}+\frac{c^{-2}}{2}\sqrt{-\gamma_{(0)}}\gamma^\alpha_{(2)\alpha}+\mathcal{O}(c^{-4}),
\end{align}
along with \eqref{e2.8} one finds
\begin{align}
\label{e2.37}
&L_B =-\frac{c^2}{2}\sqrt{-\gamma_{(0)}}\gamma^{\alpha \beta}_{(0)}T^{(0)}_{\bar{a}\bar{b}}(x)\partial_\alpha x^{\bar{a}}\partial_\beta x^{\bar{b}}-\frac{1}{2}\sqrt{-\gamma_{(0)}}\gamma^{\alpha \beta}_{(0)}\Big[T^{(0)}_{\bar{a}\bar{b}}(x)\Big(\partial_\alpha x^{\bar{a}}\partial_\beta y^{\bar{b}} \nonumber\\
&+\partial_\alpha y^{\bar{a}}\partial_\beta x^{\bar{b}} \Big)+T^{(-2)}_{\bar{a}\bar{b}}(x,y)\partial_\alpha x^{\bar{a}}\partial_\beta x^{\bar{b}}\Big]-\frac{1}{2}\sqrt{-\gamma_{(0)}}\Big(\frac{1}{2}\gamma^{\alpha \beta}_{(0)}\gamma^\lambda_{(2)\lambda}\nonumber\\
&-\gamma^{\alpha \beta}_{(2)}  \Big)T^{(0)}_{\bar{a}\bar{b}}(x)\partial_\alpha x^{\bar{a}}\partial_\beta x^{\bar{b}}-\frac{1}{2}\sqrt{-\gamma_{(0)}}\gamma^{\alpha \beta}_{(0)}\Big[ \Pi_{\bar{a}\bar{b}}^{(0)}(x)\partial_\alpha x^{\bar{a}}\partial_\beta x^{\bar{b}}\nonumber\\
&+(f(x^{\bar{a}})H^{(I)}_{\hat{\mu}\hat{\nu}}(x)+H^{(II)}_{\hat{\mu}\hat{\nu}}(x)) \partial_\alpha x^{\hat{\mu}}\partial_\beta x^{\hat{\nu}}\Big]+\mathcal{O}(c^{-2})\nonumber\\
&= c^2 L^{(2)}_B + L_B^{(0)}+\mathcal{O}(c^{-2}).
\end{align}

Before we proceed further, a few comments are in order. Notice that, the LO theory solely depends upon the fields $ x^M $ as proposed in the expansion \eqref{e2.8}. On the other hand, the sigma model at NLO encodes both the fields $ x^M $ and $ y^M $.  Therefore, to summarise, the equations of motion at LO fixes the LO fields $ x^M $ and these solutions are used further to determine the dynamics of $ y^M $ fields at NLO.
\section{Fermionic sector}
\label{sec3}
The $ \kappa $- fixed \cite{Drukker:2000ep} fermionic Lagrangian on $ AdS_5 \times S^5 $ could be expressed as
\begin{align}
\label{e2.38}
&L_F= -2i \Big( \sqrt{-\gamma}\gamma^{\alpha \beta}\bar{\vartheta}\varrho_\alpha \mathcal{D}_\beta \vartheta +\frac{i}{2L}\varepsilon^{\alpha \beta}\bar{\vartheta}\varrho_{\alpha}\Gamma_\ast \varrho_\beta  \vartheta\Big),\\
&\mathcal{D}_\alpha = \partial_\alpha +\frac{1}{4}\omega_\alpha^{\bar{A}\bar{B}}\Gamma_{\bar{A}\bar{B}}.
\end{align}

We define the projections of $ AdS_5 \times S^5 $ gamma matrices and the spin connection \cite{Hernandez:2004kr}
\begin{align}
\label{e2.32}
\varrho_\alpha = \partial_\alpha X^M E_M~^{\bar{A}}\Gamma_{\bar{A}}~;~\omega_\alpha^{\bar{A}\bar{B}}=\partial_\alpha X^M \omega_M^{\bar{A}\bar{B}},
\end{align}
together with the chirality operator along $AdS_5$ \cite{Bellucci:2005vq} $ \Gamma_{\ast}=i \Gamma_{01234} $ such that, $ \Gamma^2_{\ast}=1 $. 

Considering a NR expansion for the world-sheet spinors\footnote{The expansion \eqref{e2.35} is identical to the variation of the fermionic fields namely the dilatino and the gravitino in the NR limit of $ \mathcal{N}=1 $ supergravity in ten dimensions \cite{Bergshoeff:2021tfn}. The LO and NLO coefficients in the expansions of \cite{Bergshoeff:2021tfn} are identified as the fields of the NR minimal supergravity in ten diemsions. Likewise, the LO and NLO coefficients in \eqref{e2.35} are the fermionic excitations of the sigma model in the NR limit of GS strings.} \cite{Gomis:2005pg}
\begin{align}
\label{e2.35}
\vartheta = c^{1/2}\chi_{(1)} +c^{-1/2}\chi_{(2)}
\end{align}
and collecting all the pieces \eqref{e2.43}, \eqref{e2.58} and \eqref{e2.59} together, one can schematically $ 1/c $ expand the fermionic Lagrangian \eqref{e2.38} as\footnote{See, Appendix \ref{appena}- Appendix \ref{appenc} for details of the derivation.}
\begin{align}
\label{e2.60}
L_F = c^2 L_F^{(2)} + cL_F^{(1)} + L_F^{(0)}+\mathcal{O}(c^{-1}).
\end{align}

Here, $L_F^{(2)}$ is  the fermionic Lagrangian at LO and so on. The individual terms in the expansion \eqref{e2.60} may be expressed as
\begin{align}
&L_F^{(2)} =-2i\Big[  \sqrt{-\gamma_{(0)}}\gamma^{\alpha \beta}_{(0)}\Big( \bar{\chi}_{(1)}\varrho^{(1)}_{\alpha}\partial_\beta \chi_{(1)}+\frac{1}{4}\bar{\chi}_{(1)}\varrho_\alpha^{(1)}\Lambda_\beta^{(0)}\chi_{(1)}\Big)\nonumber\\
&+\frac{i}{2}\varepsilon^{\alpha \beta}\bar{\chi}_{(1)}\varrho^{(1)}_\alpha \Gamma_\ast  \varrho^{(1)}_\beta \chi_{(1)} \Big],\\
&L_F^{(1)} =-2i \Big[  \sqrt{-\gamma_{(0)}}\gamma^{\alpha \beta}_{(0)} \Big(  \bar{\chi}_{(1)}\varrho^{(0)}_{\alpha}\partial_\beta \chi_{(1)} + \bar{\chi}_{(1)}\varrho^{(1)}_{\alpha}\partial_\beta \chi_{(2)}+\bar{\chi}_{(2)}\varrho^{(1)}_{\alpha}\partial_\beta \chi_{(1)} \Big) \nonumber\\
&+\frac{1}{4}\sqrt{-\gamma_{(0)}}\gamma^{\alpha \beta}_{(0)}\Big( \bar{\chi}_{(1)}\varrho_\alpha^{(1)}\Lambda_\beta^{(0)}\chi_{(2)}+\bar{\chi}_{(1)}\varrho_\alpha^{(0)}\Lambda_\beta^{(0)}\chi_{(1)}+\bar{\chi}_{(2)}\varrho_\alpha^{(1)}\Lambda_\beta^{(0)}\chi_{(1)} \Big)  \nonumber\\
&+\frac{i}{2}\varepsilon^{\alpha \beta} \Big( \bar{\chi}_{(1)}\varrho^{(1)}_\alpha \Gamma_\ast  \varrho^{(1)}_\beta \chi_{(2)} +\bar{\chi}_{(1)}\varrho^{(1)}_\alpha \Gamma_\ast  \varrho^{(0)}_\beta \chi_{(1)}+\bar{\chi}_{(1)}\varrho^{(0)}_\alpha \Gamma_\ast  \varrho^{(1)}_\beta \chi_{(1)}\nonumber\\
&+\bar{\chi}_{(2)}\varrho^{(1)}_\alpha \Gamma_\ast  \varrho^{(1)}_\beta \chi_{(1)}\Big) \Big],\\
&L_F^{(0)} =-2i \Big[ \sqrt{-\gamma_{(0)}}\gamma^{\alpha \beta}_{(0)} \Big(\bar{\chi}_{(1)}\varrho^{(-1)}_{\alpha}\partial_\beta \chi_{(1)}+\bar{\chi}_{(1)}\varrho^{(0)}_{\alpha}\partial_\beta \chi_{(2)} +\bar{\chi}_{(2)}\varrho^{(0)}_{\alpha}\partial_\beta \chi_{(1)} \nonumber\\&+\bar{\chi}_{(2)}\varrho^{(1)}_{\alpha}\partial_\beta \chi_{(2)}\Big)+\sqrt{-\gamma_{(0)}}\Big(\frac{1}{2}\gamma^{\alpha \beta}_{(0)}\gamma^\lambda_{(2)\lambda}-\gamma^{\alpha \beta}_{(2)}  \Big)\bar{\chi}_{(1)}\varrho^{(1)}_{\alpha}\partial_\beta \chi_{(1)} \nonumber\\
&+\frac{1}{4}\sqrt{-\gamma_{(0)}}\gamma^{\alpha \beta}_{(0)}\Big(\bar{\chi}_{(1)}\varrho_\alpha^{(1)}\Lambda_\beta^{(-2)}\chi_{(1)}+\bar{\chi}_{(1)}\varrho_\alpha^{(0)}\Lambda_\beta^{(0)}\chi_{(2)}
\nonumber\\&+\bar{\chi}_{(1)}\varrho_\alpha^{(-1)}\Lambda_\beta^{(0)}\chi_{(1)}+\bar{\chi}_{(2)}\varrho_\alpha^{(1)}\Lambda_\beta^{(0)}\chi_{(2)}
+\bar{\chi}_{(2)}\varrho_\alpha^{(0)}\Lambda_\beta^{(0)}\chi_{(1)}\Big)\nonumber\\
&+\sqrt{-\gamma_{(0)}}\Big(\frac{1}{2}\gamma^{\alpha \beta}_{(0)}\gamma^\lambda_{(2)\lambda}-\gamma^{\alpha \beta}_{(2)}  \Big)\bar{\chi}_{(1)}\varrho_\alpha^{(1)}\Lambda_\beta^{(0)}\chi_{(1)}+\frac{i}{2}\varepsilon^{\alpha \beta} \Big(\bar{\chi}_{(1)}\varrho^{(1)}_\alpha \Gamma_\ast  \varrho^{(0)}_\beta \chi_{(1)} \nonumber\\
&+\bar{\chi}_{(1)}\varrho^{(1)}_\alpha \Gamma_\ast  \varrho^{(-1)}_\beta \chi_{(1)} +\bar{\chi}_{(1)}\varrho^{(0)}_\alpha \Gamma_\ast  \varrho^{(1)}_\beta \chi_{(2)}  + \bar{\chi}_{(1)}\varrho^{(0)}_\alpha \Gamma_\ast  \varrho^{(0)}_\beta \chi_{(1)} \nonumber\\
&+\bar{\chi}_{(1)}\varrho^{(-1)}_\alpha \Gamma_\ast  \varrho^{(1)}_\beta \chi_{(1)}+\bar{\chi}_{(2)}\varrho^{(1)}_\alpha \Gamma_\ast  \varrho^{(1)}_\beta \chi_{(2)} + \bar{\chi}_{(2)}\varrho^{(1)}_\alpha \Gamma_\ast  \varrho^{(0)}_\beta \chi_{(1)}\nonumber\\
&+\bar{\chi}_{(2)}\varrho^{(0)}_\alpha \Gamma_\ast  \varrho^{(1)}_\beta \chi_{(1)}   \Big)\Big].
\end{align}

Here, $ \varrho^{(n)}_{\alpha} $ and $ \Lambda^{(n)}_\alpha $ are the \emph{composite} excitations of the world-sheet theory those are built out of bosonic degrees of freedom namely the background SNC data. Like in the bosonic sector, we find an identical structure for the fermionic counterpart as well. One could see that the LO sigma solely depends upon fermions ($ \chi_{(1)} $) at LO, as given in the expansion \eqref{e2.35}. The solution to these fields therefore fixes the dynamics of the fermionic counterpart at LO. A detailed look further reveals that these LO fermions are also coupled with LO bosonic fields ($ x^M $) by virtue of \eqref{e3.15}. Clearly, the fermions ($ \chi_{(2)} $) at NLO could be solved provided one solves the LO bosonic ($ x^M $) as well as fermionic excitations ($ \chi_{(1)} $).

Combining \eqref{e2.37} and \eqref{e2.60}, we finally obtain a systematic string $ 1/c $ expansion for the full GS superstring action on $ AdS_5 \times S^5 $ can be schematically expressed as
\begin{align}
S_{NR}= - cT_{NR}\int d^2 \sigma \Big( L_{NR-LO}+c^{-1}L_{NR-NLO}+c^{-2}L_{NR-NNLO}+\cdots \Big),
\end{align}
where we introduce the NR string tension as $ T_{NR}=c T_{rel} $. Here, $T_{rel}$ is the string tension in the GS description. 

Below, we explore the LO Lagrangian density in detail and discuss its associated symmetries. We outline the generic structure of the NLO Lagrangian towards the end and add on some comments regarding its properties which deserve further investigations in future.
\section{Lagrangian at LO}
\label{sec4}
We first note down the leading order (LO) Lagrangian density which is the simplest one to deal with. The LO Lagrangian density could be expressed in detail as
\begin{align}
\label{e2.65}
&L_{NR-LO}=\frac{1}{2}\sqrt{-\gamma_{(0)}}\gamma^{\alpha \beta}_{(0)}\eta_{AB}\tau_\alpha~^A(x)\tau_\beta~^B (x)+2i\sqrt{-\gamma_{(0)}}\gamma^{\alpha \beta}_{(0)}\bar{\chi}_{(1)}\tau_\alpha~^A (x)\Gamma_A\hat{D}_{\beta}\chi_{(1)}\nonumber\\
&-\varepsilon^{\alpha \beta}\bar{\chi}_{(1)}\tau_\alpha~^A (x)\Gamma_A\Gamma_\ast  \tau_\beta~^B (x)\Gamma_B \chi_{(1)},
\end{align}
where we define, $ \tau_\alpha~^A (x)= \tau_{\bar{a}}~^A(x) \partial_\alpha x^{\bar{a}}$ together with $ \hat{D}_{\beta}=\partial_\beta + \frac{1}{4}\Lambda^{(0)}_\beta $ which could be thought of as being the analogue of the usual gauge covariant derivative for its NR counterpart so that the theory preserves the local scale invariance at LO.

The resulting equations of motion for bosons and fermions could be listed as follows:
\begin{align}
&\partial_\alpha \Big( \sqrt{-\gamma_{(0)}}\gamma^{\alpha \beta}_{(0)}T^{(0)}_{\bar{a}\bar{b}}(x)\partial_\beta x^{\bar{b}} \Big)+2i \partial_\alpha \Big[  \sqrt{-\gamma_{(0)}}\gamma^{\alpha \beta}_{(0)} \bar{\chi}_{(1)}\tau_{\bar{a}}~^A (x)\Gamma_A \hat{D}_\beta \chi_{(1)}\nonumber\\
&+ i \varepsilon^{\alpha \beta}\bar{\chi}_{(1)}\tau_{\bar{a}}~^A (x)\Gamma_A \Gamma_\ast \tau_\beta~^B (x) \Gamma_B \chi_{(1)}\Big]-\frac{i}{2} \sqrt{-\gamma_{(0)}}\gamma^{\alpha \beta}_{(0)}\bar{\chi}_{(1)}\tau_\alpha~^A (x)\Gamma_A\frac{\delta \Lambda^{(0)}_\beta}{\delta x^{\bar{a}}}\chi_{(1)}=0,\\
&\sqrt{-\gamma_{(0)}}\gamma^{\alpha \beta}_{(0)}\tau_\alpha~^A(x) \Gamma_A\hat{D}_{\beta}\chi_{(1)}+\frac{i}{2}\varepsilon^{\alpha \beta}\tau_\alpha~^A(x) \Gamma_A \Gamma_\ast \tau_\beta~^B(x) \Gamma_B \chi_{(1)}=0,
\label{e4.3}
\end{align}
which could be splitted into different components considering $ x^{\bar{a}}=\lbrace t^{(0)} , \rho^{(0)} \rbrace$ and using the background data those were obtained in \eqref{2.27}-\eqref{2.31}. As emphasised before, at LO one solves longitudinal fluctuations for the bosonic excitation and the leading order fermionic fields in the expansion \eqref{e2.35}. These solutions are then used at NLO and NNLO to solve remaining fluctuations. 

The 2d stress tensor associated with the world-sheet theory could be worked out as
\begin{align}
\label{e2.68}
&\mathcal{T}^{(0)}_{\alpha \beta}(x)=-\frac{2}{ \sqrt{-\gamma_{(0)}}}\frac{\delta L_{NR-LO}}{\delta \gamma^{\alpha \beta}_{(0)}}\nonumber\\
&=\frac{1}{2}\gamma_{(0)\alpha \beta}\gamma^{\lambda \kappa}_{(0)}\Big( \eta_{AB}\tau_\lambda~^A(x)\tau_\kappa~^B (x) +4i \bar{\chi}_{(1)}\tau_\lambda~^A(x)\Gamma_A \hat{D}_\kappa \chi_{(1)}\Big)\nonumber\\
&- \eta_{AB}\tau_\alpha~^A(x)\tau_\beta~^B (x) -4i \bar{\chi}_{(1)}\tau_\alpha~^A(x) \Gamma_A \hat{D}_\beta \chi_{(1)},
\end{align}
which is clearly traceless, $ \mathcal{T}^{\alpha(0)}_{\alpha}=0 $ reflecting the Weyl invariance of the NR theory at LO.

The Lagrangian \eqref{e2.65} is manifestly invariant under the world-sheet diffeomorphism. Combining the Weyl invariance along with the world-sheet diffeomorphism, one could therefore argue in favour of a gauge fixing of the world-sheet metric at LO namely
\begin{align}
\gamma_{\alpha \beta(0)}=\eta_{\alpha \beta},
\end{align}
where $ \eta_{\alpha \beta} $ is the 2d Lorentzian metric on the world-sheet.

This simplifies the LO theory \eqref{e2.65} as
\begin{align}
\label{e4.6}
&L_{NR-LO}=\frac{1}{2}\eta^{\alpha \beta}\tau_{\alpha \beta}(x)+2i\eta^{\alpha \beta}\bar{\chi}_{(1)}\tau_\alpha~^A (x)\Gamma_A\hat{D}_{\beta}\chi_{(1)}\nonumber\\
&-\varepsilon^{\alpha \beta}\bar{\chi}_{(1)}\tau_\alpha~^A (x)\Gamma_A\Gamma_\ast  \tau_\beta~^B (x)\Gamma_B \chi_{(1)},
\end{align}
where we define, $ \tau_{\alpha \beta}(x)= \eta_{AB}\tau_\alpha~^A(x)\tau_\beta~^B (x)$.
\subsection{Galilean boost}
To explore the Galilean boost symmetries of the LO theory \eqref{e4.6}, one has to introduce the world-sheet pull-back of the transverse metric and its inverse
\begin{align}
H_{\alpha \beta}=\tilde{e}_\alpha~^I \tilde{e}_\beta~^J \delta_{IJ}~;~H^{\alpha \beta}=\tilde{e}^\alpha~_I \tilde{e}^\beta~_J \delta^{IJ}
\end{align}
where the pull-back of the transverse vielbeins satisfy the normalisation
\begin{align}
\tilde{e}^\alpha~_J \tilde{e}_\alpha~^I = \delta_J^I.
\end{align}

In what follows, we impose the following transformation rules on the longitudinal vielbeins under the action of the Galilean boost \cite{Roychowdhury:2022est}
\begin{align}
\delta_G \tau_\alpha~^A =\tau_\alpha~^B G_{\gamma B}H^{\beta \gamma}\tau_\beta~^A,
\end{align}
where the Galilean boost generator is defined as \cite{Blair:2021waq}-\cite{Roychowdhury:2022est}
\begin{align}
G_\alpha~^A = \tilde{e}_\alpha~^I G_I~^A.
\end{align}

A straightforward computation reveals that under the action of the Galilean boost
\begin{align}
\delta_G L^{(B)}_{NR-LO}=\frac{1}{2}\delta_G \Big( \eta^{\alpha \beta}\tau_{\alpha \beta} \Big)=\eta^{\alpha \beta}\tau_\alpha~^C G_{\gamma C}H^{\tilde{\beta}\gamma}\tau_{\tilde{\beta}\beta}=0.
\end{align}

On the other hand, variation of the fermionic sector yields the following expression
\begin{align}
&\delta_G L^{(F)}_{NR-LO}=2i \Big(\delta_G \bar{\chi}_{(1)}\tau^\alpha~_A \Gamma^A \hat{D}_\alpha \chi_{(1)}+\bar{\chi}_{(1)}\tau^\alpha~_A \Gamma^A \delta_G\hat{D}_\alpha \chi_{(1)}\nonumber\\
&+\bar{\chi}_{(1)}\tau^\alpha~_A G_\gamma~^A H^{\beta \gamma}\tau_\beta~^B \Gamma_B \hat{D}_\alpha \chi_{(1)}\Big)\nonumber\\&-
\varepsilon^{\alpha \beta}\Big( \delta_G \bar{\chi}_{(1)}\tau_\alpha~^A \Gamma_A \Gamma_\ast \tau_\beta~^B \Gamma_B \chi_{(1)}+\bar{\chi}_{(1)}\tau_\alpha~^A \Gamma_A \Gamma_\ast \tau_\beta~^B \Gamma_B\delta_G \chi_{(1)}\nonumber\\
&+ 2\bar{\chi}_{(1)}\tau_\alpha~^A G_{\gamma A}H^{\tilde{\beta} \gamma}\tau_{\tilde{\beta}}~^C \Gamma_C \Gamma_\ast \tau_\beta~^B \Gamma_B \chi_{(1)} \Big),
\end{align}
which has to vanish in order to preserve the boost invariance of the LO theory \eqref{e4.6}.

This leads to the following set of transformation rules under the Galilean boost
\begin{align}
&\tau_\beta~^B \Gamma_A \Gamma_B \delta_G \chi_{(1)}=-2 G_{\gamma A}\tau_{\alpha}~^CH^{\alpha \gamma} \tau_\beta~^B\Gamma_C  \Gamma_B \chi_{(1)},\\
&\Gamma^A \delta_G\hat{D}_\alpha \chi_{(1)}=-G_\gamma~^A H^{\beta \gamma}\tau_\beta~^B \Gamma_B \hat{D}_\alpha \chi_{(1)},
\end{align}
while the coefficient of the third variation term $ \delta_\Lambda \bar{\chi}_{(1)} $ vanishes identically by virtue of the fermion equation of motion \eqref{e4.3}, which is therefore \emph{on-shell}.
\subsection{Dilatations}
In order to explore the dilatation invariance of the LO theory \eqref{e4.6}, we propose a local scaling of the longitudinal vielbein as \cite{Blair:2021waq}-\cite{Roychowdhury:2022est} and the gradient of world-sheet scalar \cite{Roychowdhury:2022est} as
\begin{align}
\label{e4.15}
\delta_\lambda \tau_{\bar{a}}~^A (x)= \lambda \tau_{\bar{a}}~^A (x)~;~\delta_\lambda (\partial_\alpha x^{\bar{a}})=-\lambda \partial_\alpha x^{\bar{a}}.
\end{align}

Using \eqref{e4.15}, it is trivial to show
\begin{align}
\delta_\lambda \tau_\alpha~^A =0.
\end{align}

Therefore, the variation for the bosonic sector vanishes identically
\begin{align}
\delta_\lambda L^{(B)}_{NR-LO}=0.
\end{align}

For the fermionic sector, one finds under dilatation
\begin{align}
\label{e4.18}
&\delta_\lambda L^{(F)}_{NR-LO}=2i \bar{\chi}_{(1)}\tau^\alpha~_A \Gamma^A \delta_\lambda\hat{D}_\alpha \chi_{(1)}-
\varepsilon^{\alpha \beta}\bar{\chi}_{(1)}\tau_\alpha~^A \Gamma_A \Gamma_\ast \tau_\beta~^B \Gamma_B\delta_\lambda \chi_{(1)}\nonumber\\
&+\delta_\lambda \bar{\chi}_{(1)}\Big(2i \tau^\alpha~_A \Gamma^A \hat{D}_\alpha \chi_{(1)}-\varepsilon^{\alpha \beta} \tau_\alpha~^A \Gamma_A \Gamma_\ast \tau_\beta~^B \Gamma_B \chi_{(1)} \Big).
\end{align}

Clearly, the first two terms in \eqref{e4.18} vanishes under a local scaling of the form
\begin{align}
\delta_\lambda\hat{D}_\alpha \chi_{(1)} = -\frac{i\lambda}{2}\varepsilon_{\alpha \beta}\Gamma_\ast \tau^\beta~_B \Gamma^B \chi_{(1)}~;~\delta_\lambda \chi_{(1)} = \lambda \chi_{(1)},
\end{align}
while the last term vanishes identically by virtue of \eqref{e4.3}.
\subsection{Supersymmetry}
We introduce supersymmetry transformations for the world-sheet fields at LO as
\begin{align}
\label{e4.20}
\delta_\zeta \chi_{(1)}=\zeta~;~ \delta_\zeta \tau_\alpha~^A = i  \partial_\alpha\bar{\chi}_{(1)}\Gamma^A \zeta~;~\delta_\zeta \bar{\chi}_{(1)}=\bar{\zeta},
\end{align}
where $ \zeta = \zeta (\sigma^\alpha) $ is the \emph{local} supersymmetry parameter on the world-sheet.

Using \eqref{e4.20}, for the bosonic Lagrangian one finds
\begin{align}
\delta_\zeta L^{(B)}_{NR-LO}=i \eta^{\alpha \beta}\tau_\beta~^A \partial_\alpha \bar{\chi}_{(1)}\Gamma^B \zeta \eta_{AB}.
\end{align}

For the fermionic sector, on the other hand, one finds
\begin{align}
&\delta_\zeta L^{(F)}_{NR-LO}=2i\eta^{\alpha \beta}\bar{\zeta}\tau_\alpha~^A \Gamma^B\hat{D}_{\beta}\chi_{(1)}\eta_{AB}+2 i \eta^{\alpha \beta}\bar{\chi}_{(1)}\tau_\alpha~^A \Gamma^B \delta_\zeta \partial_\beta \chi_{(1)}\eta_{AB}\nonumber\\
&+\frac{i}{2}\eta^{\alpha \beta}\bar{\chi}_{(1)}\tau_\alpha~^A \Gamma^B \Lambda^{(0)}_\beta \zeta \eta_{AB}-\varepsilon^{\alpha \beta}\bar{\zeta}\tau_\alpha~^A \Gamma^B\Gamma_\ast  \tau_\beta~^C \Gamma^D \chi_{(1)}\eta_{AB}\eta_{CD}\nonumber\\
&-\varepsilon^{\alpha \beta}\bar{\chi}_{(1)}\tau_\alpha~^A \Gamma^B\Gamma_\ast  \tau_\beta~^C \Gamma^D \zeta \eta_{AB}\eta_{CD}+ \mathcal{O}(\chi^3),
\end{align}
where $  \Lambda_\alpha^{(0)} $ is a (composite) bosonic excitation \eqref{e3.15} of the LO theory.

Using the on-shell condition \eqref{e4.3}, and imposing supersymmetry transformation 
\begin{align}
\label{69}
\delta_\zeta \partial_\alpha \chi_{(1)}=-\frac{1}{4}\Big(\Lambda_\alpha^{(0)}+2i \varepsilon_{\alpha \gamma}\eta^{\beta \gamma}\tau_\beta~^A \Gamma_\ast \Gamma_A \Big)\zeta,
\end{align}
one finds that upto quadratic order in fermions, the supersymmetry variation yields
\begin{align}
\label{e4.24}
&i\delta_\zeta L_{NR-LO}=\bar{\chi}_{(1)}\partial_\alpha j^\alpha-\partial_\alpha v^\alpha,\\
&v^\alpha =\bar{\chi}_{(1)} j^\alpha ~;~  j^\alpha=\eta^{\alpha \beta}\tau_\beta~^A\Gamma^B \zeta \eta_{AB}.
\end{align}

Clearly, the variation \eqref{e4.24} boils down into a boundary term subjected to a vanishing divergence of the form
\begin{align}
\label{e4.26}
\partial_\alpha j^\alpha = \partial_\alpha (\eta^{\alpha \beta}\tau_\beta~^A\Gamma^B \zeta \eta_{AB})=0.
\end{align}

Considering a global transformation, the above condition \eqref{e4.26} leads to
\begin{align}
\eta^{\alpha \beta}\partial_\alpha \tau_\beta~^A =0.
\end{align}
\subsection{A special case: Flat gauge}
We now choose to work with the \emph{flat} gauge which is defined by the following gauge choice
\begin{align}
\label{e4.28}
\tau_{\bar{a}}^0 = \delta^t_{\bar{a}}~;~\tau_{\bar{a}}^1 = \delta^u_{\bar{a}}~;~e_{\hat{\mu}}~^I = \delta_{\hat{\mu}}^I,
\end{align}
where we identify $ \lbrace x^t, x^u \rbrace $ as the longitudinal axes of the flat NL target space.

Using \eqref{e4.28}, one can further simplify the LO Lagrangian \eqref{e4.6} as
\begin{align}
\label{e4.29}
&L_{NR-LO}=-\frac{1}{2}\eta^{\alpha \beta}(\partial_\alpha x^t \partial_\beta x^t -\partial_\alpha x^u \partial_\beta x^u)-2i \eta^{\alpha \beta}\bar{\chi}_{(1)}\partial_\alpha x^t \Gamma^0 \partial_\beta \chi_{(1)}\nonumber\\
&+ 2i \eta^{\alpha \beta}\bar{\chi}_{(1)} \Gamma^1\partial_\alpha \chi_{(1)}\partial_\beta x^u+2\varepsilon^{\alpha \beta}\partial_\alpha x^u  \bar{\chi}_{(1)}\Gamma^1 \Gamma_\ast \Gamma^0 \chi_{(1)}\partial_\beta x^t.
\end{align}

To simplify the dynamics, we choose a string embedding of the following form
\begin{align}
\label{4.30}
x^t = \sigma^0 ~;~x^u = \sigma^1,
\end{align}
which leads to the LO Lagrangian density 
\begin{align}
L_{NR-LO}=1+2i (\bar{\chi}_{(1)}\Gamma^0 \dot{\chi}_{(1)}+\bar{\chi}_{(1)}\Gamma^1 \chi'_{(1)})-2\bar{\chi}_{(1)}\Gamma^1 \Gamma_\ast \Gamma^0 \chi_{(1)}.
\end{align}

Varying $ \bar{\chi}_{(1)} $, we obtain the corresponding equation of motion
\begin{align}
\label{e4.32}
\Gamma^0 \dot{\chi}_{(1)}+\Gamma^1 \chi'_{(1)}=-i\Gamma^1 \Gamma_\ast \Gamma^0 \chi_{(1)}.
\end{align}

Imposing only the time dependence for the world-sheet spinor, one finds 
\begin{align}
\chi_{(1)}(\sigma^0)=e^{-i\Gamma^1 \Gamma_\ast \sigma^0}\chi_{(1)}.
\end{align}

\paragraph{Longitudinal spatial T-duality.} Given the above Lagrangian \eqref{e4.29}, it is now possible to perform a T- duality following the original prescription of \cite{Rocek:1991ps}. To perform a T-duality, we compactify the longitudinal coordinate $ x^u $, which exhibits a global symmetry of the form
\begin{align}
\label{e4.30}
x^u \rightarrow x^u + \upsilon,
\end{align}
where $ \upsilon $ is a constant.

The associated Noether current turns out to be
\begin{align}
J^\alpha = \eta^{\alpha \beta}\partial_\beta x^u + 2i \eta^{\alpha \beta}\bar{\chi}_{(1)}\Gamma^1 \partial_\beta \chi_{(1)}+2\varepsilon^{\alpha \beta} \partial_\beta x^t \bar{\chi}_{(1)}\Gamma^1 \Gamma_\ast \Gamma^0 \chi_{(1)}.
\end{align}

Next, we gauge the global symmetry \eqref{e4.30} by promoting $ \upsilon = \upsilon (\sigma^\alpha) $ as a local parameter on the world-sheet and associating it with a gauge transformation of the form
\begin{align}
\delta \mathcal{A}_\alpha = \partial_\alpha \upsilon.
\end{align}

In order to preserve the gauge invariance of the LO theory, one has to implement the following replacement
\begin{align}
\partial_\alpha x^u \rightarrow \partial_\alpha x^u - \mathcal{A}_\alpha,
\end{align}
which leads to the following T-dual Lagrangian density of the form
\begin{align}
\label{e4.34}
&\tilde{L}_{NR-LO}=-\frac{1}{2}\partial^\alpha x^t \partial_\alpha x^t +\frac{1}{2}(\partial^\alpha x^u - \mathcal{A}^\alpha)(\partial_\alpha x^u - \mathcal{A}_\alpha)\nonumber\\
&-2i \partial^\alpha x^t \bar{\chi}_{(1)}\Gamma^0 \partial_\alpha \chi_{(1)}+2i (\partial^\alpha x^u - \mathcal{A}^\alpha)\bar{\chi}_{(1)}\Gamma^1 \partial_\alpha \chi_{(1)}\nonumber\\
&+ 2 \varepsilon^{\alpha \beta}(\partial_\alpha x^u - \mathcal{A}_\alpha) \partial_\beta x^t \bar{\chi}_{(1)}\Gamma^1 \Gamma_\ast \Gamma^0 \chi_{(1)}-\frac{\lambda}{2}  \varepsilon^{\alpha \beta}(\partial_\alpha \mathcal{A}_\beta -\partial_\beta \mathcal{A}_\alpha),
\end{align}
where the Lagrange multiplier ($ \lambda $) ensures the \emph{flatness} of the world-sheet gauge connection.

To proceed further, we introduce the following field redefinition
\begin{align}
\tilde{\mathcal{A}}_\alpha = \mathcal{A}_\alpha - \partial_\alpha x^u ,
\end{align}
which re-expresses the T-dual Lagrangian \eqref{e4.34} as
\begin{align}
\label{e4.36}
&\tilde{L}_{NR-LO}=-\frac{1}{2}\partial^\alpha x^t \partial_\alpha x^t + \frac{1}{2}\tilde{\mathcal{A}}^\alpha \tilde{\mathcal{A}}_\alpha -2i \partial^\alpha x^t \bar{\chi}_{(1)}\Gamma^0 \partial_\alpha \chi_{(1)}\nonumber\\
&-2i \tilde{\mathcal{A}}^\alpha \bar{\chi}_{(1)}\Gamma^1 \partial_\alpha \chi_{(1)}- 2 \varepsilon^{\alpha \beta} \tilde{\mathcal{A}}_\alpha\partial_\beta x^t \bar{\chi}_{(1)}\Gamma^1 \Gamma_\ast \Gamma^0 \chi_{(1)}\nonumber\\
&-\frac{\lambda}{2} \varepsilon^{\alpha \beta}(\partial_\alpha \tilde{\mathcal{A}}_\beta -\partial_\beta \tilde{\mathcal{A}}_\alpha).
\end{align}

The Lagrange multiplier ($ \lambda $) imposes the following constraint
\begin{align}
\partial_\alpha \tilde{\mathcal{A}}_\beta -\partial_\beta \tilde{\mathcal{A}}_\alpha =0,
\end{align}
which is satisfied by choosing
\begin{align}
\label{e4.38}
\tilde{\mathcal{A}}_\alpha = \partial_\alpha \xi.
\end{align}

Substituting \eqref{e4.38} into \eqref{e4.36} and comparing with \eqref{e4.29}, one finds
\begin{align}
\xi = -x^u.
\end{align}

One can re-express \eqref{e4.36} as
\begin{align}
\label{e4.40}
&\tilde{L}_{NR-LO}=-\frac{1}{2}\partial^\alpha x^t \partial_\alpha x^t + \frac{1}{2}\tilde{\mathcal{A}}^\alpha \tilde{\mathcal{A}}_\alpha -2i \partial^\alpha x^t \bar{\chi}_{(1)}\Gamma^0 \partial_\alpha \chi_{(1)}\nonumber\\
&-2i \tilde{\mathcal{A}}^\alpha \bar{\chi}_{(1)}\Gamma^1 \partial_\alpha \chi_{(1)}- 2 \varepsilon^{\alpha \beta} \tilde{\mathcal{A}}_\alpha\partial_\beta x^t \bar{\chi}_{(1)}\Gamma^1 \Gamma_\ast \Gamma^0 \chi_{(1)}- \varepsilon^{\alpha \beta}\tilde{\mathcal{A}}_\alpha \partial_\beta \lambda,
\end{align}
where $ \tilde{\mathcal{A}}_\alpha $ acts as a Lagrange multiplier which can be integrated out to yield
\begin{align}
\label{e4.41}
\tilde{\mathcal{A}}_\alpha =2i \bar{\chi}_{(1)}\Gamma^1 \partial_\alpha \chi_{(1)}+2 \eta_{\alpha \gamma}\varepsilon^{\gamma \beta}\partial_\beta x^t \bar{\chi}_{(1)}\Gamma^1 \Gamma_\ast \Gamma^0 \chi_{(1)}+\eta_{ \alpha \gamma}\varepsilon^{\gamma \beta}\partial_\beta \lambda.
\end{align}

Substituting \eqref{e4.41} back into \eqref{e4.40} and setting $ \lambda = \tilde{x}^{\tilde{u}} $ (dual coordinate), we finally note down the T-dual Lagrangian upto quadratic order in fermions as
\begin{align}
&\tilde{L}_{NR-LO}=-\frac{1}{2}\eta^{\alpha \beta}\partial_\alpha x^t \partial_\beta x^t +\frac{1}{2}\tilde{\eta}^{\alpha \beta}\partial_\alpha \tilde{x}^{\tilde{u}}\partial_{\beta}\tilde{x}^{\tilde{u}}-2i \eta^{\alpha \beta} \partial_\alpha x^t \bar{\chi}_{(1)}\Gamma^0 \partial_\beta \chi_{(1)}\nonumber\\
&- 2i \varepsilon^{\alpha \beta}\bar{\chi}_{(1)}\Gamma^1 \partial_\alpha \chi_{(1)}\partial_\beta \tilde{x}^{\tilde{u}}+2 \tilde{\eta}^{\alpha\beta}\partial_{\alpha}\tilde{x}^{\tilde{u}}\bar{\chi}_{(1)}\Gamma^1 \Gamma_\ast \Gamma^0 \chi_{(1)}\partial_\beta x^t,
\end{align}
where we denote, $ \tilde{\eta}^{\alpha \beta}=-\eta_{\gamma \tilde{\gamma}}\varepsilon^{\gamma \alpha}\varepsilon^{\tilde{\gamma} \beta} $.
\section{Concluding remarks}
\label{sec5}
Let us first summarize the key results of this paper. The present paper is an attempt to understand the non-relativistic (NR) expansion of Green-Schwarz (GS) superstrings on $ AdS_5 \times S^5 $ where we restrict ourselves up to quadratic order in the fermions. The NR reduction yields a sigma model Lagrangian that can be expressed as a string $ 1/c $ expansion where we retain ourdelves up to $ \mathcal{O}(c^{-2}) $. We explore the LO theory in detail including its local as well as global symmetries. As a special case, we choose the flat gauge and explore the associated (longitudinal) T-duality properties of the LO Lagrangian.

Before we conclude, it would be nice to discuss some of the properties of the NR theory at NLO. As a matter of fact, the NLO theory has a sole contribution that comes from the fermionic sector \eqref{e2.60} of the original relativistic model
\begin{align}
\label{nlo}
&(-2i)^{-1}L_{NR-NLO}=\sqrt{-\gamma_{(0)}}\gamma^{\alpha \beta}_{(0)} \Big( \sqrt{f(x^{\bar{a}})} \bar{\chi}_{(1)}e_{\alpha}~^I \Gamma_I \partial_\beta \chi_{(1)} + \bar{\chi}_{(1)}\tau_{\alpha}~^A \Gamma_A \partial_\beta \chi_{(2)}\nonumber\\&
+\bar{\chi}_{(2)}\tau_{\alpha}~^A \Gamma_A\partial_\beta \chi_{(1)} \Big) +\frac{1}{4}\sqrt{-\gamma_{(0)}}\gamma^{\alpha \beta}_{(0)}\Big( \bar{\chi}_{(1)}\tau_{\alpha}~^A \Gamma_A\Lambda_\beta^{(0)}\chi_{(2)}\nonumber\\
&+\sqrt{f(x^{\bar{a}})}\bar{\chi}_{(1)}e_{\alpha}~^I \Gamma_I\Lambda_\beta^{(0)}\chi_{(1)}+\bar{\chi}_{(2)}\tau_{\alpha}~^A \Gamma_A\Lambda_\beta^{(0)}\chi_{(1)} \Big)  \nonumber\\
&+\frac{i}{2}\varepsilon^{\alpha \beta} \Big( \bar{\chi}_{(1)}\tau_{\alpha}~^A \Gamma_A \Gamma_\ast  \tau_{\beta}~^B \Gamma_B \chi_{(2)} +\sqrt{f(x^{\bar{a}})}\bar{\chi}_{(1)}\tau_{\alpha}~^A \Gamma_A \Gamma_\ast  e_{\beta}~^I \Gamma_I \chi_{(1)}\nonumber\\
&+\sqrt{f(x^{\bar{a}})}\bar{\chi}_{(1)}e_{\alpha}~^I \Gamma_I \Gamma_\ast  \tau_{\beta}~^A \Gamma_A \chi_{(1)}+\bar{\chi}_{(2)}\tau_{\alpha}~^A \Gamma_A \Gamma_\ast  \tau_{\beta}~^B \Gamma_B \chi_{(1)}\Big).
\end{align}

A straightforward computation reveals that the stress tensor yields a vanishing trace
\begin{align}
\mathcal{T}_{\alpha}^{\alpha(1)}=0,
\end{align}
which therefore preserves the Weyl invariance of the sigma model at NLO.

Clearly, the theory at NLO is a bit more involved and one could in principle solve for NLO fermion fluctuations by knowing the solutions at previous order. The same applies when we discuss the symmetries of the theory namely the transformation rules under global/ local symmetries are governed by the transformation properties of the world-sheet excitations at LO. The supersymmetry transformations of fermions and the bosonic fields at LO would dictate the supersymmetry transformation rules for fermionic fields at NLO. This is due to the fact that at NLO, we have contributions coming both from $ \chi_{(1)} $ and $ \chi_{(2)} $. While $ \chi_{(1)} $ is the field at LO in the expansion \eqref{e2.35}, $ \chi_{(2)} $ is the field at NLO. Also notice that $ \chi_{(2)} $ is coupled with LO bosonic excitation $ \tau_\alpha~^A $. Combining all these pieces together, one might consider a variation of the NLO Lagrangian \eqref{nlo} under the supersymmetry transformation which can be schematically expressed as
\begin{align}
\label{95}
&\delta_\zeta L_{NR-NLO}=\eta^{\alpha \beta}\Big( \delta_\zeta\bar{\chi}_{(1)}\tau_{\alpha}~^A \Gamma_A \partial_\beta \chi_{(2)}+\bar{\chi}_{(1)}\delta_\zeta\tau_{\alpha}~^A \Gamma_A \partial_\beta \chi_{(2)}+\bar{\chi}_{(1)}\tau_{\alpha}~^A \Gamma_A \delta_\zeta\partial_\beta \chi_{(2)} \nonumber\\
&+\delta_\zeta\bar{\chi}_{(2)}\tau_{\alpha}~^A \Gamma_A\partial_\beta \chi_{(1)}+\bar{\chi}_{(2)}\delta_\zeta\tau_{\alpha}~^A \Gamma_A\partial_\beta \chi_{(1)}+\bar{\chi}_{(2)}\tau_{\alpha}~^A \Gamma_A\delta_\zeta\partial_\beta \chi_{(1)} \Big)+\cdots
\end{align}
where we choose to work with the conformal gauge for simplicity. Clearly, the supersymmetry transformation rules for the NLO field $ \chi_{(2)} $ and its derivative could be fixed by setting the variation \eqref{95} equal to zero and using the set of transformation rules \eqref{e4.20} and \eqref{69} that are obtained at LO. Similar structure persists beyond NLO. For example, world-sheet fields and the supersymmetry variation of their derivatives at NNLO are governed by the supersymmetry variation of LO and NLO fields at previous orders. However, the theory becomes more involved as one continues to probe higher order in the $ 1/c $ expansion.

Before we conclude, a couple of points are worth pressing here. 

$\bullet$ It is now quite compelling in trying to understand the dynamics of the background fields from the perspective of a beta function calculation \cite{Gallegos:2019icg} and pin down the low energy effective action that might have also been expanded in a systematic $1/c$ expansion. From the perspective of the sigma model, it is expected that the LO effective action would preserve some amount of background supersymmetry. Yet another interesting direction would be to explore the scattering amplitude and estimate the full quantum spectrum. Since the sigma model becomes highly nontrivial, it would indeed be a challenge to perform the above calculations beyond LO.

$\bullet$ The major difference between the present analysis and that of \cite{Gomis:2005pg} rests on the fact that the analysis of \cite{Gomis:2005pg} had been performed by fixing the $\kappa$- symmetry gauge which eventually gets rid of the leading order coefficient in the fermionic expansion \eqref{e2.35}. In other words, one sets $\chi_{(1)}=0$. As a result, the fermionic expansion eventually starts with the $\chi_{(2)}$ field which acts like a LO contribution in the $\kappa$- symmetric gauge. This is precisely reflected in the finite contribution as found in \cite{Gomis:2005pg} 
\begin{align}
\label{96}
\mathcal{L}_{fin}\sim \sqrt{-h}h^{\alpha \beta}\eta_{\bar{a}\bar{b}}(\text{v}_{(\alpha}~^{\bar{a}} \text{v}_{\beta)}~^{\bar{b}} +\text{v}_{(\alpha}~^{\bar{a}} \bar{\theta}_+ \gamma^{\bar{b}}(-\sigma_2) D_{\beta)} \theta_+ )+\sqrt{-h}h^{\alpha \beta}\delta_{\hat{\mu}\hat{\nu}}\partial_\alpha X^{\hat{\mu}} \partial_\beta X^{\hat{\nu}}
\end{align}
where we make the following identifications, $\theta_+ =\chi_{(2)}$ along with $\eta_{\bar{a}\bar{b}}= \eta_{AB}\tau_{\bar{a}}~^A \tau_{\bar{b}}~^B$. The first term in the parenthesis on the r.h.s. of \eqref{96} is therefore structurally identical to the LO action \eqref{e2.65} except for the fact that the fermionic contribution starts with a $\mathcal{O}(1/\sqrt{c})$ term which is an artefact of the $\kappa$- symmetric gauge as emphasized above. Finally, a closer look at transverse fields $ X^{\hat{\mu}} $ reveals a trivial dynamics in the conformal gauge, where they are completely decoupled from the longitudinal modes $ \text{v}_{\alpha}~^{\bar{a}} \sim \partial_\alpha X^{\bar{a}} $ at LO.

\paragraph{$  \kappa$- symmetry.} Following \cite{Gomis:2005pg}, it is possible to carry out a similar analysis for $ \kappa $- symmetry in the present NR framework. Below we out line the key steps, which starts with the basic definition that sets the rule under $ \kappa $- transformation
\begin{align}
\label{97}
\delta_{\kappa}\vartheta = \frac{1}{2}(1- \Gamma_\kappa)\kappa ~;~ \delta_{\kappa}X^M =0
\end{align}
where we expand the $ \kappa $- parameter as \cite{Gomis:2005pg}
\begin{align}
\kappa = \sqrt{c}\kappa_{(1)}+\frac{1}{\sqrt{c}}\kappa_{(2)}.
\end{align}

The $\Gamma_\kappa$ matrix, on the other hand, is given by \cite{Gomis:2005pg}
\begin{align}
\label{98}
\Gamma_\kappa = \frac{1}{2 \sqrt{\det G_{\alpha \beta}}}\varepsilon^{\alpha \beta}\slashed{\Gamma}_{\alpha}\slashed{\Gamma}_{\beta} \tau_3.
\end{align}

A straightforward computation reveals
\begin{align}
\label{100}
\sqrt{\det G_{\alpha \beta}}=c^2 \sqrt{\det T^{(0)}_{\alpha \beta}(x)}(1+\mathcal{O}(c^{-2}))
\end{align}
where we denote, $ T^{(0)}_{\alpha \beta}(x) = T^{(0)}_{\bar{a}\bar{b}}(x)\partial_\alpha x^{\bar{a}}\partial_\beta x^{\bar{b}} $.

On a similar note, the projection of the gamma matrices could be expanded as
\begin{align}
\label{101}
\slashed{\Gamma}_{\alpha}= c \tau_\alpha~^A \Gamma_A +e_\alpha~^I \Gamma_I+ \mathcal{O}(c^{-1}).
\end{align}

Using \eqref{100}-\eqref{101}, one finds an expansion of the following form
\begin{align}
\label{102}
\Gamma_\kappa =\frac{\varepsilon^{\alpha \beta} \slashed{\tau}_\alpha \slashed{\tau}_\beta}{2\sqrt{\det T^{(0)}_{\alpha \beta}(x)}} \tau_{3} +\mathcal{O}(c^{-1})\equiv \Gamma_\kappa^{(0)}+\mathcal{O}(c^{-1})
\end{align}
where we denote, $ \slashed{\tau}_\alpha= \tau_\alpha~^A \Gamma_A$.

Substituting \eqref{102} back into \eqref{97} and using \eqref{98} one can obtain $ \kappa $- symmetry transformation rules for fermions at different order in the expansion \eqref{e2.35}. For example, for fermion field at LO in the expansion \eqref{e2.35}, one finds
\begin{align}
\delta_{\kappa}\chi_{(1)} = \frac{1}{2}(1- \Gamma^{(0)}_\kappa)\kappa_{(1)}
\end{align}
and so on. Substituting the $ \kappa $- variations into \eqref{e2.65}, and thereby demanding the invariance of the LO action, one could fix the $ \kappa $- transformation rule for fermion derivative $ \hat{D}_\alpha \chi_{(1)} $.

$ \bullet $ Finally, it is worth looking at the AdS/CFT implications of the NR sigma model obtained in this paper. One possible way to realise this is to take a world-sheet Galilean limit following the recent prescription of \cite{Bidussi:2023rfs}. This is achieved by rescaling the longitudinal vielbein, $ \tau_u~^1 \rightarrow \frac{1}{\tilde{c}} \tau_u~^1$ and thereby taking the $ \tilde{c}\rightarrow \infty $ limit. Here, $ \tilde{c} $ is the speed of light along the compact (longitudinal) isometry $ x^u $. The resulting procedure would give rise to what one should identify as the Galilean GS superstrings, which could in principle be related to the Spin-Matrix limit \cite{Harmark:2014mpa} of $ \mathcal{N}=4 $ SYM, following the arguments of \cite{Harmark:2018cdl}, \cite{Bidussi:2023rfs}.
\paragraph{Acknowledgements.}
The author is indebted to the authorities of IIT Roorkee for their unconditional support towards researches in basic sciences. The author also acknowledges The Royal Society, UK for financial assistance. The author is also indebted to the Department of Physics, Swansea University, UK where some part of this work was done. The author would also like to acknowledge S. Prem Kumar and Carlos Nunez for several illuminating discussions.
\appendix
\section{Expansion of gamma matrices}
\label{appena}
Let us first consider a $ 1/c $ expansion of 2d projected gamma matrices
\begin{align}
\label{e2.34}
&\varrho_\alpha = \partial_\alpha X^M E_M~^{\bar{A}}\Gamma_{\bar{A}}\nonumber\\
&= \partial_\alpha X^{\bar{a}} E_{\bar{a}}~^{A}\Gamma_{A}+ \partial_\alpha X^{\hat{\mu}} E_{\hat{\mu}}~^{I}\Gamma_{I}\nonumber\\
&=c \tau_{\bar{a}}~^A (x)\partial_\alpha x^{\bar{a}}\Gamma_A+\sqrt{f(x^{\bar{a}})}\partial_\alpha x^{\hat{\mu}}\tilde{e}_{\hat{\mu}}~^I(x)\Gamma_I +c^{-1}\Big[\partial_\alpha x^{\bar{a}}y^M \partial_M \tau_{\bar{a}}~^A (x)\nonumber\\
&+ \partial_\alpha x^{\bar{a}}m_{\bar{a}}~^A (x)+\partial_\alpha y^{\bar{a}}\tau_{\bar{a}}~^A(x)  \Big]\Gamma_A +c^{-2}\Big[\sqrt{f(x^{\bar{a}})}\partial_\alpha x^{\hat{\mu}}y^M \partial_M \tilde{e}_{\hat{\mu}}~^I (x)\nonumber\\
&+ \sqrt{f(x^{\bar{a}})}\Big(\frac{1}{2}\partial_\alpha x^{\hat{\mu}}f^{-1}(x^{\bar{a}})y^M \partial_M f (x^{\bar{a}})+\partial_\alpha y^{\hat{\mu}}  \Big)\tilde{e}_{\hat{\mu}}~^I(x)  \Big]\Gamma_I + \mathcal{O}(c^{-3})\nonumber\\
&= c \varrho_\alpha^{(1)}+\varrho_\alpha^{(0)}+c^{-1}\varrho_\alpha^{(-1)}+c^{-2}\varrho_\alpha^{(-2)}+\mathcal{O}(c^{-3}).
\end{align}

Combining \eqref{e2.34} and \eqref{e2.35}, one finally obtains
\begin{align}
\label{e2.43}
 &\sqrt{-\gamma}\gamma^{\alpha \beta}\bar{\vartheta}\varrho_\alpha \partial_\beta \vartheta = c^2 \sqrt{-\gamma_{(0)}}\gamma^{\alpha \beta}_{(0)} \bar{\chi}_{(1)}\varrho^{(1)}_{\alpha}\partial_\beta \chi_{(1)}+c  \sqrt{-\gamma_{(0)}}\gamma^{\alpha \beta}_{(0)} \Big[  \bar{\chi}_{(1)}\varrho^{(0)}_{\alpha}\partial_\beta \chi_{(1)} \nonumber\\
 &+ \bar{\chi}_{(1)}\varrho^{(1)}_{\alpha}\partial_\beta \chi_{(2)}+\bar{\chi}_{(2)}\varrho^{(1)}_{\alpha}\partial_\beta \chi_{(1)} \Big]+ \sqrt{-\gamma_{(0)}}\gamma^{\alpha \beta}_{(0)} \Big[\bar{\chi}_{(1)}\varrho^{(-1)}_{\alpha}\partial_\beta \chi_{(1)}\nonumber\\&+\bar{\chi}_{(1)}\varrho^{(0)}_{\alpha}\partial_\beta \chi_{(2)} +\bar{\chi}_{(2)}\varrho^{(0)}_{\alpha}\partial_\beta \chi_{(1)} +\bar{\chi}_{(2)}\varrho^{(1)}_{\alpha}\partial_\beta \chi_{(2)}\Big]+\sqrt{-\gamma_{(0)}}\Big(\frac{1}{2}\gamma^{\alpha \beta}_{(0)}\gamma^\lambda_{(2)\lambda}\nonumber\\
&-\gamma^{\alpha \beta}_{(2)}  \Big)\bar{\chi}_{(1)}\varrho^{(1)}_{\alpha}\partial_\beta \chi_{(1)}+c^{-1}\Big[ \sqrt{-\gamma_{(0)}}\gamma^{\alpha \beta}_{(0)}\Big( \bar{\chi}_{(1)}\varrho^{(-1)}_{\alpha}\partial_\beta \chi_{(2)}+\bar{\chi}_{(2)}\varrho^{(-1)}_{\alpha}\partial_\beta \chi_{(1)}\nonumber\\
&+\bar{\chi}_{(2)}\varrho^{(0)}_{\alpha}\partial_\beta \chi_{(2)} +\bar{\chi}_{(1)}\varrho^{(-2)}_{\alpha}\partial_\beta \chi_{(1)}\Big)+\sqrt{-\gamma_{(0)}}\Big(\frac{1}{2}\gamma^{\alpha \beta}_{(0)}\gamma^\lambda_{(2)\lambda}-\gamma^{\alpha \beta}_{(2)}  \Big)\nonumber\\
 &\Big(  \bar{\chi}_{(1)}\varrho^{(0)}_{\alpha}\partial_\beta \chi_{(1)} + \bar{\chi}_{(1)}\varrho^{(1)}_{\alpha}\partial_\beta \chi_{(2)}+\bar{\chi}_{(2)}\varrho^{(1)}_{\alpha}\partial_\beta \chi_{(1)} \Big) \Big]+\mathcal{O}(c^{-2}).
\end{align}
\section{Expansion of the spin connection}
Next, we consider the NR expansion of the spin connection on the world-sheet. To this end, we first note down the target space spin connection
\begin{align}
\label{e2.39}
\omega_M^{\bar{A}\bar{B}} = \eta^{\bar{B}\bar{C}}\omega_M~^{\bar{A}}~_{\bar{C}}=\eta^{\bar{B}\bar{C}} E_P~^{\bar{A}}E^N~_{\bar{C}}\hat{\Gamma}^P_{M N}-\eta^{\bar{B}\bar{C}}E^N~_{\bar{C}}\partial_M E_{N}~^{\bar{A}},
\end{align}
where $ \eta^{\bar{A}\bar{B}} $ is the flat ten dimensional Lorentzian metric together with $ \hat{\Gamma}^A_{BC} $ being the usual Christoffel connections in ten dimensions.

Using \eqref{e2.39} and \eqref{e2.3}-\eqref{2.11}, we find
\begin{align}
&\omega_M^{\bar{A}\bar{B}}\Gamma_{\bar{A}\bar{B}}= \eta^{BC}\Big[ \tau_{\bar{a}}~^A (x)\tau^{\bar{b}}~_C (x)+ c^{-2}\Big(  y^M \partial_M \tau_{\bar{a}}~^A (x)\tau^{\bar{b}}~_C (x)\nonumber\\&+\tau_{\bar{a}}~^A (x)y^M \partial_M\tau^{\bar{b}}~_C (x)+\tau_{\bar{a}}~^A (x)m^{\bar{b}}~_C (x)+m_{\bar{a}}~^A (x)\tau^{\bar{b}}~_C (x) \Big)  \Big]\hat{\Gamma}^{\bar{a}}_{M \bar{b}}\Gamma_{AB}\nonumber\\
&+\delta^{JK}\Big[ \tilde{e}_{\hat{\mu}}~^I (x)\tilde{e}^{\hat{\nu}}~_K (x) +c^{-2}\Big( y^N \partial_N\tilde{e}_{\hat{\mu}}~^I (x)\tilde{e}^{\hat{\nu}}~_K (x)\nonumber\\
& + \tilde{e}_{\hat{\mu}}~^I (x)y^N \partial_N\tilde{e}^{\hat{\nu}}~_K (x)  \Big)\Big]\hat{\Gamma}^{\hat{\mu}}_{M \hat{\nu}}\Gamma_{IJ}-\eta^{BC}\tau^{\bar{a}}~_C (x)\partial_M \tau_{\bar{a}}~^A (x)\Gamma_{AB}\nonumber\\
&-c^{-2}\eta^{BC}\Big(\tau^{\bar{a}}~_C (x)\partial_M m_{\bar{a}}~^A (x)+m^{\bar{a}}~_C (x)\partial_M \tau_{\bar{a}}~^A (x)\nonumber\\
&+ y^N \partial_N \tau^{\bar{a}}~_C (x)\partial_M \tau_{\bar{a}}~^A (x)+\tau^{\bar{a}}~_C (x)y^N \partial_N \partial_M \tau_{\bar{a}}~^A (x)\Big)\Gamma_{AB}\nonumber\\
&- \delta^{JK}\tilde{e}^{\hat{\mu}}~_K (x)\partial_M \tilde{e}_{\hat{\mu}}~^I (x)\Gamma_{IJ}-c^{-2} \delta^{JK}\Big[y^N \partial_N \tilde{e}^{\hat{\mu}}~_K (x)\partial_M \tilde{e}_{\hat{\mu}}~^I (x)\nonumber\\
&+\tilde{e}^{\hat{\mu}}~_K (x)y^N \partial_N\partial_M \tilde{e}_{\hat{\mu}}~^I (x) \Big]\Gamma_{IJ},
\end{align}
where $ A,B,C =(0,1)$ are the flat two dimensional Lorentzian indices and $ I,J,K $ are the flat indices associated with the eight dimensional transverse space of the NL manifold.

\paragraph{Expansion of $ \hat{\Gamma}^P_{MN} $ :} Consider the general expression for Christoffel connection over a general relativity background
\begin{align}
\hat{\Gamma}^P_{MN} =\frac{1}{2}G^{PK}(\partial_M G_{NK}+\partial_N G_{KM}-\partial_K G_{MN}).
\end{align}

Following \eqref{2.9}, the general form of the metric could be read off as
\begin{align}
&G_{MN}=c^2 T_{MN}(X)+\Pi_{MN}(X)+f(x^{\bar{a}})H_{MN}(X)+ \mathcal{O}(c^{-2}).
\end{align}

The inverse metric is defined as
\begin{align}
\label{e2.48}
G^{MN}=f^{-1}(x^{\bar{a}})H^{MN}(X)+c^{-2}T^{MN}(X),
\end{align}
such that following the orthonoramality condition $ G_{MN}G^{NP}=\delta^P_M $, one finds
\begin{align}
T_{MN}H^{NP}=0=\Pi_{MN}H^{NP}~;~ T_{MN}T^{NP}+H_{MN}H^{NP}=\delta^P_M .
\end{align}

Using \eqref{e2.48}, one finds the following expansion for the Christoffel connection
\begin{align}
\label{e2.50}
&\hat{\Gamma}^P_{MN} = \frac{c^2}{2}f^{-1}H^{PK}\Big( \partial_M T_{NK}+\partial_N T_{MK}-\partial_K T_{MN} \Big)+\frac{f^{-1}}{2}H^{PK}\Big[ \partial_M \Pi_{NK}\nonumber\\
&+ \partial_M (f H_{NK}) +\partial_N \Pi_{MK}+ \partial_N (f H_{MK})  -\partial_K \Pi_{MN}- \partial_K (f H_{MN}) \Big]\nonumber\\
&+\frac{1}{2}T^{PK}\Big(  \partial_M T_{NK}+\partial_N T_{MK}-\partial_K T_{MN} \Big)+\frac{c^{-2}}{2}T^{PK}\Big[ \partial_M \Pi_{NK}\nonumber\\
&+ \partial_M (f H_{NK}) +\partial_N \Pi_{MK}+ \partial_N (f H_{MK})  -\partial_K \Pi_{MN}- \partial_K (f H_{MN}) \Big].
\end{align}

Using \eqref{e2.50}, one can finally express the world-sheet spin connection as
\begin{align}
\label{e2.51}
\omega_\alpha^{\bar{A}\bar{B}}\Gamma_{\bar{A}\bar{B}}=\partial_\alpha X^M \omega_M^{\bar{A}\bar{B}}\Gamma_{\bar{A}\bar{B}}=\Lambda^{(0)}_\alpha + c^{-2}\Lambda^{(-2)}_\alpha+ \mathcal{O}(c^{-4}),
\end{align}
where we identify the above functions as
\begin{align}
\label{e3.15}
& \Lambda^{(0)}_\alpha = \eta^{BC}\tau_{\bar{a}}~^A (x)\tau^{\bar{b}}~_C (x)\hat{\Gamma}^{(0)\bar{a}}_{M \bar{b}}(x)\Gamma_{AB}\partial_\alpha x^M +\delta^{JK} \tilde{e}_{\hat{\mu}}~^I (x)\tilde{e}^{\hat{\nu}}~_K (x)\hat{\Gamma}^{(0)\hat{\mu}}_{M \hat{\nu}}(x)\Gamma_{IJ}\partial_\alpha x^M \nonumber\\
&-\eta^{BC}\tau^{\bar{a}}~_C (x)\partial_M \tau_{\bar{a}}~^A (x)\Gamma_{AB}\partial_\alpha x^M- \delta^{JK}\tilde{e}^{\hat{\mu}}~_K (x)\partial_M \tilde{e}_{\hat{\mu}}~^I (x)\Gamma_{IJ} \partial_\alpha x^M,\\
& \Lambda^{(-2)}_\alpha =\eta^{BC}\tau_{\bar{a}}~^A (x)\tau^{\bar{b}}~_C (x)\hat{\Gamma}^{(-2)\bar{a}}_{M \bar{b}}(x,y)\Gamma_{AB}\partial_\alpha x^M +\eta^{BC}\Big(  y^M \partial_M \tau_{\bar{a}}~^A (x)\tau^{\bar{b}}~_C (x)\nonumber\\&+\tau_{\bar{a}}~^A (x)y^M \partial_M\tau^{\bar{b}}~_C (x)+\tau_{\bar{a}}~^A (x)m^{\bar{b}}~_C (x)+m_{\bar{a}}~^A (x)\tau^{\bar{b}}~_C (x) \Big) \hat{\Gamma}^{(0)\bar{a}}_{M \bar{b}}(x)\Gamma_{AB}\partial_\alpha x^M\nonumber\\
&+\eta^{BC}\tau_{\bar{a}}~^A (x)\tau^{\bar{b}}~_C (x)\hat{\Gamma}^{(0)\bar{a}}_{M \bar{b}}(x)\Gamma_{AB}\partial_\alpha y^M +\delta^{JK} \tilde{e}_{\hat{\mu}}~^I (x)\tilde{e}^{\hat{\nu}}~_K (x)\hat{\Gamma}^{(-2)\hat{\mu}}_{M \hat{\nu}}(x,y)\Gamma_{IJ}\partial_\alpha x^M \nonumber\\
&+\delta^{JK}\Big( y^N \partial_N\tilde{e}_{\hat{\mu}}~^I (x)\tilde{e}^{\hat{\nu}}~_K (x) + \tilde{e}_{\hat{\mu}}~^I (x)y^N \partial_N\tilde{e}^{\hat{\nu}}~_K (x)  \Big)\hat{\Gamma}^{(0)\hat{\mu}}_{M \hat{\nu}}(x)\Gamma_{IJ}\partial_\alpha x^M\nonumber\\
&+\delta^{JK} \tilde{e}_{\hat{\mu}}~^I (x)\tilde{e}^{\hat{\nu}}~_K (x)\hat{\Gamma}^{(0)\hat{\mu}}_{M \hat{\nu}}(x)\Gamma_{IJ}\partial_\alpha y^M-\eta^{BC}\tau^{\bar{a}}~_C (x)\partial_M \tau_{\bar{a}}~^A (x)\Gamma_{AB}\partial_\alpha y^M\nonumber\\
&-\eta^{BC}\Big(\tau^{\bar{a}}~_C (x)\partial_M m_{\bar{a}}~^A (x)+m^{\bar{a}}~_C (x)\partial_M \tau_{\bar{a}}~^A (x)\nonumber\\
&+ y^N \partial_N \tau^{\bar{a}}~_C (x)\partial_M \tau_{\bar{a}}~^A (x)+\tau^{\bar{a}}~_C (x)y^N \partial_N \partial_M \tau_{\bar{a}}~^A (x)\Big)\Gamma_{AB} \partial_\alpha x^M \nonumber\\
&- \delta^{JK}\tilde{e}^{\hat{\mu}}~_K (x)\partial_M \tilde{e}_{\hat{\mu}}~^I (x)\Gamma_{IJ} \partial_\alpha y^M- \delta^{JK}\Big[y^N \partial_N \tilde{e}^{\hat{\mu}}~_K (x)\partial_M \tilde{e}_{\hat{\mu}}~^I (x)\nonumber\\
&+\tilde{e}^{\hat{\mu}}~_K (x)y^N \partial_N\partial_M \tilde{e}_{\hat{\mu}}~^I (x) \Big]\Gamma_{IJ}\partial_\alpha x^M,
\end{align}
together with the expansions of the Christoffel connection as
\begin{align}
&\hat{\Gamma}^{(0)\bar{a}}_{M \bar{b}}=\frac{1}{2}T^{(0)\bar{a}\bar{c}}\Big(  \partial_M T^{(0)}_{\bar{b}\bar{c}}+\partial_{\bar{b}} T^{(0)}_{M\bar{c}}-\partial_{\bar{c}} T^{(0)}_{M\bar{b}} \Big),\\
&\hat{\Gamma}^{(-2)\bar{a}}_{M \bar{b}}=\frac{1}{2}T^{(-2)\bar{a}\bar{c}}\Big(  \partial_M T^{(0)}_{\bar{b}\bar{c}}+\partial_{\bar{b}} T^{(0)}_{M\bar{c}}-\partial_{\bar{c}} T^{(0)}_{M\bar{b}} \Big)\nonumber\\
&+\frac{1}{2}T^{(0)\bar{a}\bar{c}}\Big(  \partial_M T^{(-2)}_{\bar{b}\bar{c}}+\partial_{\bar{b}} T^{(-2)}_{M\bar{c}}-\partial_{\bar{c}} T^{(-2)}_{M\bar{b}} \Big)\nonumber\\
&+\frac{1}{2}T^{(0)\bar{a}\bar{c}}\Big(  \partial_M \Pi^{(0)}_{\bar{b}\bar{c}}+\partial_{\bar{b}} \Pi^{(0)}_{M\bar{c}}-\partial_{\bar{c}} \Pi^{(0)}_{M\bar{b}} \Big),
\end{align}
where the functions $ T^{(0)}(x) $, $ T^{(-2)}(x,y) $ and $ \Pi^{(0)}(x) $ are given in \eqref{e2.9} and \eqref{e2.10}.

The rest of the Christoffel connections can be expressed as
\begin{align}
&\hat{\Gamma}^{(0)\hat{\mu}}_{M \hat{\nu}}=\frac{f^{-1}}{2}H^{\hat{\mu}\hat{\lambda}}(x)\Big[\partial_M (f H_{\hat{\nu}\hat{\lambda}}(x))+\partial_{\hat{\nu}} (f H_{M\hat{\lambda}}(x)) -\partial_{\hat{\lambda}}(f H_{M \hat{\nu}})(x) \Big],\\
&\hat{\Gamma}^{(-2)\hat{\mu}}_{M \hat{\nu}}=\frac{f^{-1}}{2}y^L \partial_L H^{\hat{\mu}\hat{\lambda}}(x)\Big[\partial_M (f H_{\hat{\nu}\hat{\lambda}}(x))+\partial_{\hat{\nu}} (f H_{M\hat{\lambda}}(x)) -\partial_{\hat{\lambda}}(f H_{M \hat{\nu}}(x) )\Big]\nonumber\\
&+\frac{f^{-1}}{2}H^{\hat{\mu}\hat{\lambda}}(x)\Big[f \Big(y^L \partial_L \partial_M H_{\hat{\nu}\hat{\lambda}}(x)+y^L \partial_L \partial_{\hat{\nu}} H_{M\hat{\lambda}}(x)-y^L \partial_L \partial_{\hat{\lambda}} H_{M\hat{\nu}}(x) \Big)\nonumber\\
&+\partial_M f y^L \partial_L H_{\hat{\nu}\hat{\lambda}}(x)+\partial_{\hat{\nu}}f y^L \partial_L H_{M \hat{\lambda}}(x)-\partial_{\hat{\lambda}}f y^L \partial_L H_{M \hat{\nu}}(x) \Big].
\end{align}

Using \eqref{e2.51}, one can finally express
\begin{align}
\label{e2.58}
&\sqrt{-\gamma}\gamma^{\alpha \beta}\bar{\vartheta}\varrho_\alpha \omega_\beta^{\bar{A}\bar{B}}\Gamma_{\bar{A}\bar{B}} \vartheta=c^2 \sqrt{-\gamma_{(0)}}\gamma^{\alpha \beta}_{(0)}\bar{\chi}_{(1)}\varrho_\alpha^{(1)}\Lambda_\beta^{(0)}\chi_{(1)}+c \sqrt{-\gamma_{(0)}}\gamma^{\alpha \beta}_{(0)}\Big[ \bar{\chi}_{(1)}\varrho_\alpha^{(1)}\Lambda_\beta^{(0)}\chi_{(2)}\nonumber\\
&+\bar{\chi}_{(1)}\varrho_\alpha^{(0)}\Lambda_\beta^{(0)}\chi_{(1)}+\bar{\chi}_{(2)}\varrho_\alpha^{(1)}\Lambda_\beta^{(0)}\chi_{(1)} \Big]+\sqrt{-\gamma_{(0)}}\gamma^{\alpha \beta}_{(0)}\Big[\bar{\chi}_{(1)}\varrho_\alpha^{(1)}\Lambda_\beta^{(-2)}\chi_{(1)}\nonumber\\&+\bar{\chi}_{(1)}\varrho_\alpha^{(0)}\Lambda_\beta^{(0)}\chi_{(2)}
+\bar{\chi}_{(1)}\varrho_\alpha^{(-1)}\Lambda_\beta^{(0)}\chi_{(1)}+\bar{\chi}_{(2)}\varrho_\alpha^{(1)}\Lambda_\beta^{(0)}\chi_{(2)}
+\bar{\chi}_{(2)}\varrho_\alpha^{(0)}\Lambda_\beta^{(0)}\chi_{(1)}\Big]\nonumber\\
&+\sqrt{-\gamma_{(0)}}\Big(\frac{1}{2}\gamma^{\alpha \beta}_{(0)}\gamma^\lambda_{(2)\lambda}-\gamma^{\alpha \beta}_{(2)}  \Big)\bar{\chi}_{(1)}\varrho_\alpha^{(1)}\Lambda_\beta^{(0)}\chi_{(1)}+c^{-1}\Big[\sqrt{-\gamma_{(0)}}\gamma^{\alpha \beta}_{(0)}\Big[ \bar{\chi}_{(1)}\varrho_\alpha^{(1)}\Lambda_\beta^{(-2)}\chi_{(2)}\nonumber\\
&+\bar{\chi}_{(1)}\varrho_\alpha^{(0)}\Lambda_\beta^{(-2)}\chi_{(1)}+\bar{\chi}_{(1)}\varrho_\alpha^{(-1)}\Lambda_\beta^{(0)}\chi_{(2)}
+\bar{\chi}_{(1)}\varrho_\alpha^{(-2)}\Lambda_\beta^{(0)}\chi_{(1)}+\bar{\chi}_{(2)}\varrho_\alpha^{(1)}\Lambda_\beta^{(-2)}\chi_{(1)}\nonumber\\
&+ \bar{\chi}_{(2)}\varrho_\alpha^{(0)}\Lambda_\beta^{(0)}\chi_{(2)}+\bar{\chi}_{(2)}\varrho_\alpha^{(-1)}\Lambda_\beta^{(0)}\chi_{(1)}\Big]+\sqrt{-\gamma_{(0)}}\Big(\frac{1}{2}\gamma^{\alpha \beta}_{(0)}\gamma^\lambda_{(2)\lambda}-\gamma^{\alpha \beta}_{(2)}  \Big)\nonumber\\
&\Big( \bar{\chi}_{(1)}\varrho_\alpha^{(1)}\Lambda_\beta^{(0)}\chi_{(2)}+ \bar{\chi}_{(1)}\varrho_\alpha^{(0)}\Lambda_\beta^{(0)}\chi_{(1)}+\bar{\chi}_{(2)}\varrho_\alpha^{(1)}\Lambda_\beta^{(0)}\chi_{(1)}\Big)\Big]+\mathcal{O}(c^{-2}).
\end{align}
\section{Expansion of the mass term}
\label{appenc}
Finally, we expand the mass term \cite{Hernandez:2004kr}-\cite{Bellucci:2005vq} in \eqref{e2.38} which is sourced due to the background five form flux ($ F_5 $). A straightforward computation reveals
\begin{align}
\label{e2.59}
&\frac{i}{2L}\varepsilon^{\alpha \beta}\bar{\vartheta}\varrho_{\alpha}\Gamma_\ast \varrho_\beta  \vartheta =c^2\frac{i}{2}\varepsilon^{\alpha \beta}\bar{\chi}_{(1)}\varrho^{(1)}_\alpha \Gamma_\ast  \varrho^{(1)}_\beta \chi_{(1)}+c\frac{i}{2}\varepsilon^{\alpha \beta} \Big[ \bar{\chi}_{(1)}\varrho^{(1)}_\alpha \Gamma_\ast  \varrho^{(1)}_\beta \chi_{(2)} \nonumber\\
&+\bar{\chi}_{(1)}\varrho^{(1)}_\alpha \Gamma_\ast  \varrho^{(0)}_\beta \chi_{(1)}+\bar{\chi}_{(1)}\varrho^{(0)}_\alpha \Gamma_\ast  \varrho^{(1)}_\beta \chi_{(1)}+\bar{\chi}_{(2)}\varrho^{(1)}_\alpha \Gamma_\ast  \varrho^{(1)}_\beta \chi_{(1)}\Big]\nonumber\\
&+\frac{i}{2}\varepsilon^{\alpha \beta} \Big[\bar{\chi}_{(1)}\varrho^{(1)}_\alpha \Gamma_\ast  \varrho^{(0)}_\beta \chi_{(1)} +\bar{\chi}_{(1)}\varrho^{(1)}_\alpha \Gamma_\ast  \varrho^{(-1)}_\beta \chi_{(1)} +\bar{\chi}_{(1)}\varrho^{(0)}_\alpha \Gamma_\ast  \varrho^{(1)}_\beta \chi_{(2)}  \nonumber\\
&+ \bar{\chi}_{(1)}\varrho^{(0)}_\alpha \Gamma_\ast  \varrho^{(0)}_\beta \chi_{(1)} +\bar{\chi}_{(1)}\varrho^{(-1)}_\alpha \Gamma_\ast  \varrho^{(1)}_\beta \chi_{(1)}+\bar{\chi}_{(2)}\varrho^{(1)}_\alpha \Gamma_\ast  \varrho^{(1)}_\beta \chi_{(2)} \nonumber\\
&+ \bar{\chi}_{(2)}\varrho^{(1)}_\alpha \Gamma_\ast  \varrho^{(0)}_\beta \chi_{(1)}+\bar{\chi}_{(2)}\varrho^{(0)}_\alpha \Gamma_\ast  \varrho^{(1)}_\beta \chi_{(1)}   \Big]+\mathcal{O}(c^{-1}),
\end{align}
where we identify $ \Gamma_\ast = \Gamma_{01}\Gamma_{I_2 I_3 I_4} $ and set $ \ell =1 $ in the final step. Here, we drop the remaining terms in the expansion as they won't be important in what follows.


\begin{thebibliography}{99}
\bibitem{Gomis:2000bd}
J.~Gomis and H.~Ooguri,
``Nonrelativistic closed string theory,''
J. Math. Phys. \textbf{42}, 3127-3151 (2001)
doi:10.1063/1.1372697
[arXiv:hep-th/0009181 [hep-th]].

\bibitem{Danielsson:2000gi}
U.~H.~Danielsson, A.~Guijosa and M.~Kruczenski,
``IIA/B, wound and wrapped,''
JHEP \textbf{10}, 020 (2000)
doi:10.1088/1126-6708/2000/10/020
[arXiv:hep-th/0009182 [hep-th]].

\bibitem{Gomis:2005pg}
J.~Gomis, J.~Gomis and K.~Kamimura,
``Non-relativistic superstrings: A New soluble sector of AdS(5) x S**5,''
JHEP \textbf{12}, 024 (2005)
doi:10.1088/1126-6708/2005/12/024
[arXiv:hep-th/0507036 [hep-th]].

\bibitem{Kim:2007pc}
B.~S.~Kim,
``Non-relativistic superstring theories,''
Phys. Rev. D \textbf{76}, 126013 (2007)
doi:10.1103/PhysRevD.76.126013
[arXiv:0710.3203 [hep-th]].

\bibitem{Gomis:2016zur}
J.~Gomis and P.~K.~Townsend,
``The Galilean Superstring,''
JHEP \textbf{02}, 105 (2017)
doi:10.1007/JHEP02(2017)105
[arXiv:1612.02759 [hep-th]].

\bibitem{Bergshoeff:2018yvt}
E.~Bergshoeff, J.~Gomis and Z.~Yan,
``Nonrelativistic String Theory and T-Duality,''
JHEP \textbf{11}, 133 (2018)
doi:10.1007/JHEP11(2018)133
[arXiv:1806.06071 [hep-th]].

\bibitem{Bergshoeff:2019pij}
E.~A.~Bergshoeff, J.~Gomis, J.~Rosseel, C.~Simsek and Z.~Yan,
``String Theory and String Newton-Cartan Geometry,''
J. Phys. A \textbf{53}, no.1, 014001 (2020)
doi:10.1088/1751-8121/ab56e9
[arXiv:1907.10668 [hep-th]].

\bibitem{Bergshoeff:2018vfn}
E.~A.~Bergshoeff, K.~T.~Grosvenor, C.~Simsek and Z.~Yan,
``An Action for Extended String Newton-Cartan Gravity,''
JHEP \textbf{01}, 178 (2019)
doi:10.1007/JHEP01(2019)178
[arXiv:1810.09387 [hep-th]].

\bibitem{Roychowdhury:2019olt}
D.~Roychowdhury,
``Nonrelativistic pulsating strings,''
JHEP \textbf{09}, 002 (2019)
doi:10.1007/JHEP09(2019)002
[arXiv:1907.00584 [hep-th]].

\bibitem{Blair:2019qwi}
C.~D.~A.~Blair,
``A worldsheet supersymmetric Newton-Cartan string,''
JHEP \textbf{10}, 266 (2019)
doi:10.1007/JHEP10(2019)266
[arXiv:1908.00074 [hep-th]].

\bibitem{Bergshoeff:2021bmc}
E.~A.~Bergshoeff, J.~Lahnsteiner, L.~Romano, J.~Rosseel and C.~\c{S}im\c{s}ek,
``A non-relativistic limit of NS-NS gravity,''
JHEP \textbf{06}, 021 (2021)
doi:10.1007/JHEP06(2021)021
[arXiv:2102.06974 [hep-th]].

\bibitem{Bergshoeff:2021tfn}
E.~A.~Bergshoeff, J.~Lahnsteiner, L.~Romano, J.~Rosseel and C.~Simsek,
``Non-relativistic ten-dimensional minimal supergravity,''
JHEP \textbf{12}, 123 (2021)
doi:10.1007/JHEP12(2021)123
[arXiv:2107.14636 [hep-th]].

\bibitem{Harmark:2017rpg}
T.~Harmark, J.~Hartong and N.~A.~Obers,
``Nonrelativistic strings and limits of the AdS/CFT correspondence,''
Phys. Rev. D \textbf{96}, no.8, 086019 (2017)
doi:10.1103/PhysRevD.96.086019
[arXiv:1705.03535 [hep-th]].

\bibitem{Harmark:2018cdl}
T.~Harmark, J.~Hartong, L.~Menculini, N.~A.~Obers and Z.~Yan,
``Strings with Non-Relativistic Conformal Symmetry and Limits of the AdS/CFT Correspondence,''
JHEP \textbf{11}, 190 (2018)
doi:10.1007/JHEP11(2018)190
[arXiv:1810.05560 [hep-th]].

\bibitem{Gallegos:2019icg}
A.~D.~Gallegos, U.~G\"ursoy and N.~Zinnato,
``Torsional Newton Cartan gravity from non-relativistic strings,''
JHEP \textbf{09}, 172 (2020)
doi:10.1007/JHEP09(2020)172
[arXiv:1906.01607 [hep-th]].

\bibitem{Hartong:2021ekg}
J.~Hartong and E.~Have,
``Nonrelativistic Expansion of Closed Bosonic Strings,''
Phys. Rev. Lett. \textbf{128}, no.2, 021602 (2022)
doi:10.1103/PhysRevLett.128.021602
[arXiv:2107.00023 [hep-th]].

\bibitem{Hartong:2022dsx}
J.~Hartong and E.~Have,
``Nonrelativistic approximations of closed bosonic string theory,''
JHEP \textbf{02}, 153 (2023)
doi:10.1007/JHEP02(2023)153
[arXiv:2211.01795 [hep-th]].

\bibitem{Harmark:2019upf}
T.~Harmark, J.~Hartong, L.~Menculini, N.~A.~Obers and G.~Oling,
``Relating non-relativistic string theories,''
JHEP \textbf{11}, 071 (2019)
doi:10.1007/JHEP11(2019)071
[arXiv:1907.01663 [hep-th]].

\bibitem{Bidussi:2021ujm}
L.~Bidussi, T.~Harmark, J.~Hartong, N.~A.~Obers and G.~Oling,
``Torsional string Newton-Cartan geometry for non-relativistic strings,''
JHEP \textbf{02}, 116 (2022)
doi:10.1007/JHEP02(2022)116
[arXiv:2107.00642 [hep-th]].

\bibitem{Yan:2021lbe}
Z.~Yan,
``Torsional deformation of nonrelativistic string theory,''
JHEP \textbf{09}, 035 (2021)
doi:10.1007/JHEP09(2021)035
[arXiv:2106.10021 [hep-th]].

\bibitem{Ebert:2021mfu}
S.~Ebert, H.~Y.~Sun and Z.~Yan,
``Dual D-brane actions in nonrelativistic string theory,''
JHEP \textbf{04}, 161 (2022)
doi:10.1007/JHEP04(2022)161
[arXiv:2112.09316 [hep-th]].

\bibitem{Blair:2021waq}
C.~D.~A.~Blair, D.~Gallegos and N.~Zinnato,
``A non-relativistic limit of M-theory and 11-dimensional membrane Newton-Cartan geometry,''
JHEP \textbf{10}, 015 (2021)
doi:10.1007/JHEP10(2021)015
[arXiv:2104.07579 [hep-th]].

\bibitem{Roychowdhury:2022est}
D.~Roychowdhury,
``Nonrelativistic expansion of M2 branes and M theory backgrounds,''
JHEP \textbf{11}, 152 (2022)
doi:10.1007/JHEP11(2022)152
[arXiv:2208.05646 [hep-th]].

\bibitem{Green:1983wt}
M.~B.~Green and J.~H.~Schwarz,
``Covariant Description of Superstrings,''
Phys. Lett. B \textbf{136}, 367-370 (1984)
doi:10.1016/0370-2693(84)92021-5

\bibitem{Green:1983sg}
M.~B.~Green and J.~H.~Schwarz,
``Properties of the Covariant Formulation of Superstring Theories,''
Nucl. Phys. B \textbf{243}, 285-306 (1984)
doi:10.1016/0550-3213(84)90030-0

\bibitem{Metsaev:1998it}
R.~R.~Metsaev and A.~A.~Tseytlin,
``Type IIB superstring action in AdS(5) x S**5 background,''
Nucl. Phys. B \textbf{533}, 109-126 (1998)
doi:10.1016/S0550-3213(98)00570-7
[arXiv:hep-th/9805028 [hep-th]].

\bibitem{Metsaev:2000yf}
R.~R.~Metsaev and A.~A.~Tseytlin,
``Superstring action in AdS(5) x S**5. Kappa symmetry light cone gauge,''
Phys. Rev. D \textbf{63}, 046002 (2001)
doi:10.1103/PhysRevD.63.046002
[arXiv:hep-th/0007036 [hep-th]].

\bibitem{Drukker:2000ep}
N.~Drukker, D.~J.~Gross and A.~A.~Tseytlin,
``Green-Schwarz string in AdS(5) x S**5: Semiclassical partition function,''
JHEP \textbf{04}, 021 (2000)
doi:10.1088/1126-6708/2000/04/021
[arXiv:hep-th/0001204 [hep-th]].

\bibitem{Hyun:2000hr}
S.~Hyun and H.~Shin,
``Supersymmetry of Green-Schwarz superstring and matrix string theory,''
Phys. Rev. D \textbf{64}, 046008 (2001)
doi:10.1103/PhysRevD.64.046008
[arXiv:hep-th/0012247 [hep-th]].

\bibitem{Hernandez:2004kr}
R.~Hernandez and E.~Lopez,
``Spin chain sigma models with fermions,''
JHEP \textbf{11}, 079 (2004)
doi:10.1088/1126-6708/2004/11/079
[arXiv:hep-th/0410022 [hep-th]].

\bibitem{Bellucci:2005vq}
S.~Bellucci, P.~Y.~Casteill and J.~F.~Morales,
``Superstring sigma models from spin chains: The SU(1,1|1) case,''
Nucl. Phys. B \textbf{729}, 163-178 (2005)
doi:10.1016/j.nuclphysb.2005.09.012
[arXiv:hep-th/0503159 [hep-th]].

\bibitem{Rocek:1991ps}
M.~Rocek and E.~P.~Verlinde,
``Duality, quotients, and currents,''
Nucl. Phys. B \textbf{373}, 630-646 (1992)
doi:10.1016/0550-3213(92)90269-H
[arXiv:hep-th/9110053 [hep-th]].

\bibitem{Bidussi:2023rfs}
L.~Bidussi, T.~Harmark, J.~Hartong, N.~A.~Obers and G.~Oling,
``Longitudinal Galilean and Carrollian limits of non-relativistic strings,''
[arXiv:2309.14467 [hep-th]].

\bibitem{Harmark:2014mpa}
T.~Harmark and M.~Orselli,
``Spin Matrix Theory: A quantum mechanical model of the AdS/CFT correspondence,''
JHEP \textbf{11}, 134 (2014)
doi:10.1007/JHEP11(2014)134
[arXiv:1409.4417 [hep-th]].
\end{thebibliography}
\end{document}